\documentclass[envcountsame,orivec]{llncs}

\usepackage{proof}
\usepackage{fullpage}
\usepackage{latexsym}
\usepackage{amssymb}

\usepackage[all]{xy}

\newcommand\enc[2]{\{#1\}_{#2}}
\newcommand\spout[3]{\bar{#1}\langle #2 \rangle.#3}
\newcommand\splet[4]{\hbox{let $\spr #1 #2 = #3$ in $#4$}}
\newcommand\spcase[4]{\hbox{case $#1$ of $\enc {#2} {#3}$ in $#4$}}
\newcommand\spr[2]{\langle #1, #2 \rangle}
\newcommand\freeNm[1]{\hbox{fn}(#1)}
\newcommand\freeRN[1]{\hbox{rn}(#1)}
\newcommand\freeFN[1]{\hbox{fn}(#1)}
\newcommand\freeRFN[1]{\hbox{rfn}(#1)}
\def\eqm{\leftrightarrow}

\newcommand\dom[1]{\hbox{dom}(#1)}

\newcommand\one[3]{#1 \stackrel{#2}{\longrightarrow} #3}
\newcommand\irred[1]{#1 \! \Downarrow}

\newcommand\defeq{\stackrel{\Delta}{=}}

\newcommand\rsubst[2]{{#1}_{\restriction {#2}}}

\def\relbar{\mathrel{\smash-}}
\def\joinrelm{\mathrel{\mkern-3mu}}
\def\tailpiece{\kern 1pt\vrule height 1ex width 0.3ex depth -.3ex}
\def\seqsym{\mathrel{\tailpiece\joinrelm\relbar}}

\def\obisim{\approx_o}

\def\abf{{\mathbf a}}
\def\bbf{{\mathbf b}}
\def\cbf{{\mathbf c}}
\def\dbf{{\mathbf d}}
\def\ebf{{\mathbf e}}
\def\kbf{{\mathbf k}}
\def\mbf{{\mathbf m}}

\begin{document}

\title{A Trace Based Bisimulation for the Spi Calculus}
\author{Alwen Tiu}
\institute{The Australian National University}

\maketitle

\begin{abstract}
A notion of open bisimulation is formulated for the spi calculus, 
an extension of the $\pi$-calculus with cryptographic primitives. 
In this formulation, open bisimulation is indexed by pairs of symbolic traces,
which represent the history of interactions between the 
environment with the pairs of processes being
checked for bisimilarity. The use of symbolic traces allows for
a symbolic treatment of bound input in bisimulation checking which
avoids quantification over input values. 
Open bisimilarity is shown to be sound with respect to testing equivalence, 
and futher, it is shown to be an equivalence relation on processes 
and a congruence relation on finite processes. 
As far as we know, this is the first formulation of open bisimulation for 
the spi calculus for which the congruence result is proved. 
\end{abstract}


\section{Introduction}

The spi-calculus \cite{abadi99ic} is an extension of the 
$\pi$-calculus~\cite{milner92icI,milner92icII}
with crytographic primitives. This extension allows one to model 
cryptographic protocols and, via a notion of observational equivalence, 
called {\em testing equivalence}, one can express security properties that a protocol satisfies. 
Testing equivalence is usually defined by quantifying the environment with which the processes
interact: roughly, to show that two processes are testing equivalent, 
one shows that the two processes exhibit the same traces under arbitrary observers. 
As in the $\pi$-calculus, bisimulation techniques have been defined to check observational 
equivalence of processes that avoids quantification over all possible observers. 
Unlike the $\pi$-calculus, in order to capture security notions such as 
secrecy, bisimulation in the spi-calculus need to take into account 
the states of the environment (e.g., public networks)
in its interaction with the processes being checked for equivalence. 
This gives rise to a more refined notion of equivalence of actions
in the definition of bisimulation. In the $\pi$-calculus,
to check whether two processes are bisimilar, one checks that an action
by a process is matched by an equivalent action by the other process,
and their continuations possess the same property. 
The differences between bisimulations for the $\pi$- and the spi-calculus 
lie in the interpretation of ``equivalent actions''; there are situations
where equivalence of actions may be interpreted as ``indistinguishable
actions'', from the perspective of an observer, which may not
be syntactically equal.

Consider the processes $P = (\nu x) \spout a {\enc b x} 0$
and $Q = (\nu x) \spout a {\enc c x} 0.$ $P$ is a process that can
output on channel $a$ a message $b$, encrypted with a fresh key $x$, and terminates,
while $Q$ outputs a message $c$ encrypted with $x$ on the same channel.
In the standard definitions of bisimulation for the $\pi$-calculus, e.g.,
late or early bisimulation ~\cite{milner92icI,milner92icII}, 
these two processes are not bisimilar since
they output (syntactically) distinct actions. In the spi-calculus, when one
is concerned only with whether an intruder (in its interaction with $P$ and $Q$)
can discover the message being encrypted, the two actions by $P$ and $Q$ are essentially
indistinguishable; the intruder does not have access to the key $x$, hence
cannot access the underlying messages.

Motivated by the above observation, different notions of bisimulation have been proposed, 
among others {\em framed bisimulation}~\cite{abadi98njc},
{\em environment-sensitive bisimulation}~\cite{boreale02sjc},
{\em hedged bisimulation}~\cite{borgstrom05mscs}, etc. 
(see ~\cite{borgstrom05mscs} for a review on these bisimulations).
All these notions of  bisimulation 
share a similarity in that they are all indexed
by some sort of structure representing the ``knowledge'' of the environment. 
This structure is called differently from one definition to another. We shall use
the rather generic term {\em observer theory}, or {\em theory} for short, to refer to 
the knowledge structure used in this paper, which is just a finite set of pairs of messages.
A theory represents the pairs of messages that are obtained through the interaction between 
the environment (observer) and the pairs of processes in the bisimulation set. 
The pairs of messages in the theory represent equivalent messages, from the point of view of 
the observer. This observer theory is then used as a theory in a deductive 
system for deducing messages (or actions) equivalence. Under this theory, equivalent
messages need not be syntactically equivalent. 


A main difficulty in bisimulation checking for spi-processes is in dealing 
with the input actions of the processes, where one needs to check that 
the processes are bisimilar for all equivalent pairs of input messages. 
One way of dealing with the infinite quantification is through
a symbolic technique
where one delays the instantiations of input values until they are needed. 
This technique has been applied to hedged bisimulation 
by Borgstr\"om et al.\cite{borgstrom04concur}. 
Their work on symbolic bisimulation for the spi-calculus
is, however, mainly concerned with obtaining a sound approximation of hedged bisimulation,
and less with studying meta-level properties of the symbolic bisimulation as an
equivalence relation. {\em Open bisimulation}~\cite{sangiorgi96acta}, on the other hand, 
makes use of the symbolic handling of input values, while at the same time maintains interesting
meta-level properties, such as being a congruence relation on processes. 
Open bisimulation 
has so far been studied for the $\pi$-calculus and its extension
to the spi-calculus has not been fully understood. There is a recent
attempt at formulating an open-style bisimulation for the spi-calculus~\cite{briais06entcs}, which
is shown to be sound with respect to hedged bisimulation. 
However, no congruence results have been obtained for this notion
of open bisimulation. We propose a different formulation of open bisimulation, 
which is inspired by hedged bisimulation. A collection of {\em up-to techniques} are defined,
and shown to be sound. These up-to techniques can be used to finitely check the bisimilarity
of processes in some cases and, more importantly, they are used to show that open bisimilarity 
is a congruence on finite spi-processes.  The latter allows for compositional reasoning
about open bisimilarity.
As far as we know, this is the first congruence result for open bisimulation for the spi calculus.

There are several novel features of our work that distinguish it from existing formulations
of bisimulation of the spi calculus. Each of these is discussed briefly below.

\subsection{Sequent calculus for observer theories}

In most formulation of bisimulation for the spi calculus, the observer's capability
in making logical inferences (e.g., deducing, from the availability of an encrypted
message $\enc M K$ and a key $K$, the message $M$) is presented as some sort of
natural deduction system. For example, suppose $\Sigma$ represents a set of
messages accumulated by an observer. Let us denote with $\Sigma \vdash M$
the fact that the observer can ``deduce $M$ from $\Sigma$''. 
Then the capability of the observer to decrypt message can be represented as the elimination rule:
$$
\infer[]
{\Sigma \vdash M}
{\Sigma \vdash \enc M K & \Sigma \vdash K}
$$
One drawback of such a representation of capability is that it is not immediately clear
how {\em proof search} for the judgment $\Sigma \vdash M$ can be done, 
since this would involve application of the rule in a bottom-up fashion, which in turn
would involve ``guessing'' a suitable key $K$. 

In this paper, we use a different representation of observer's capabilities using 
sequent calculus. The sequent calculus formulation has the advantage the the
rules are {\em local}, in the sense that, any proof of $\Sigma \vdash M$ involves
only subterms of $\Sigma$ and $M$. As it is well-known in proof theory and functional programming, 
there is a close correspondence betweent the two formalisms, e.g., the Curry-Howard 
correspondence between natural deduction and sequent calculus for intuitionistic logic. 
There is a more-or-less straightforward translation from elimination rules in natural
deduction rules to ``left-introduction'' rules in sequent calculus. The latter means that
the rules are applied to messages on the left of the turnstile $\vdash.$
For example, the above elimination rule has the corresponding left-rule in sequent calculus:
$$
\infer[]
{\Sigma, \enc M K \vdash R}
{\Sigma, \enc M K \vdash K & \Sigma, \enc M K, M, K \vdash R}
$$

For the correspondence to work, we need to show a certain transitivity property of the
sequent calculus system, that is, if $\Sigma \vdash M$ and $\Sigma, M \vdash R$ are provable,
then so is $\Sigma \vdash R.$ 
In proof theory, this result is often referred to as the {\em cut-elimination} theorem. 

Beside guaranteeing tractability of proof search, the sequent calculus formulation
of observer theory, in particular the cut elimination theorem, turns out to be useful in establishing
the metatheory of our formulation of open bisimulation. But we note that equivalent results
can be obtained using the more traditional natural deduction formulation, but perhaps with some 
extra efforts. 
Recently, sequent calculus has been used to derive decidability results
for a range of observer theories (under richer equational theories than that covered
in this paper) in a uniform way \cite{tiu08intruder}. 

\subsection{Consistency of observer theories}
A crucial part in theories of environment-sensitive bisimulation is that
of the consistency of the observer theory. Recall that an observer theory
is a set of pairs of messages, representing the history of interaction
between the observer and the pair of processes being checked for bisimilarity.
Consistency of such a theory can be roughly understood as the property of
``indistinguishability'' between the first and the second projections of the pairs.
More precisely, whatever operations one can perform on the first projections
(decrypting the messages, encrypting, testing for syntactic equality, etc.) can also be
performed on the second projections. 
A consistent theory guarantees that the induced equality on messages (or
more precisely, indistinguishability) satisfies the usual axioms of equality,
most importantly, transitivity. This in turns is used to show that the environment-sensitive
bisimulation that are parameterized upon consistent theories is an equivalence
relation.

In most previous formulations of bisimulation for the spi-calculus, the definition
of consistency is defined only on theories in a certain ``reduced form'' (see
e.g. \cite{abadi98njc,borgstrom05mscs}). 
One problem with this definition of consistency is that the reduced form is not
closed under arbitrary substitution of names. This makes it difficult to 
define the notion of consistency and reduced form for observer theories used in 
open bisimulation, since open bisimulation  involves substitution of names at arbitrary stages
in bisimulation checking, e.g., as in the original definition of
open bisimulation for the $\pi$-calculus \cite{sangiorgi96acta}.
In this paper, we define a new notion of consistency for observer theories,
which do not require the observer theories to be in reduced form. 
We then show that there is a finite (and decidable) characterisation of
consistency of any given observer theory (see Section~\ref{sec:obsv}).

\subsection{Symbolic representation of observer theories}

One difficulty in formulating open bisimulation for the spi-calculus is how to
ensure that open bisimilarity is closed under substitutions of names. Open bisimilarity
for the $\pi$-calculus is known to be not closed under arbitrary situations, so it cannot
be the case either for the spi-calculus. The question then is for what 
class of substitutions they are closed under. In the $\pi$-calculus, this class
of substitutions is defined via a notion called {\em distinction}~\cite{sangiorgi96acta}, 
which constraints the identification of certain names in the processes. 
A {\em respectful substitution}, with respect to a distinction $D$, is any substitution 
that satisfies the constraint on the distinction of names in $D$.
In the spi-calculus, input values can be arbitrary terms, not just names, 
therefore a simple notion of distinction
would not suffice. We also have to take into account the knowledge that is accumulated
by the environment in its interaction with processes. 
Consider for example the pair of processes
$
P = (\nu k) \spout a {\enc b k} {a(x).0}
$
and 
$Q = (\nu k) \spout a {\enc c k} {a(x).0}$
where $a$, $b$ and $c$ are pairwise distinct names.
Intuitively, we can see that the two processes are bisimilar, since the key $k$
is not explicitly extruded. A ``symbolic'' bisimulation game on these processes
would look something like the following diagram:
$$
\xymatrix{
P \ar[d]_{\bar a \enc b k} & \approx & Q \ar[d]^{\bar a \enc c k} \\
a(x) \ar[d]_{a x} & \approx & a(x) \ar[d]^{a x} \\
0 & \approx & 0
}
$$
where we left the input value $x$ unspecified. 
To show the soundness of this symbolic bisimulation, we have to ``concretize''
this symbolic set, by considering approriate instantiations of $x$.
Obviously, $x$ cannot be substituted by an arbitrary term, for example, it cannot
be instantiated with $k$, since this would be inconsistent with the fact that
$k$ is not explicitly extruded. We also need to take into account different
instantiations of $x$ for the continuations of $P$ and $Q$. For example,
in its interaction with $P$, the environment does not have the message $\enc c k$,
so $x$ cannot be instantiated with this term. Likewise, in its interaction with 
$Q$, it is never the case that $x$ would be instantiated with $\enc b k.$
Thus, a good notion of respectful substitutions for open bisimulation must 
respect the different knowledge of the process pairs in the bisimulation.

The symbolic representation of observer theories used in this paper is based
on Boreale's {\em symbolic traces}~\cite{boreale01icalp}. A symbolic trace is
a compact representation of a set of traces of a process, where the input values
are represented by parameters (which are essentially names). Associated with a symbolic
trace is a notion of consistency, i.e., it should be possible to instantiate the
symbolic trace to a set of concrete traces. The definition of open bisimulation
in Section~\ref{sec:open} is indexed by pairs of symbolic traces, which we call
{\em bi-traces}. A symbolic trace is essentially a list, and the position of
a particular name in the list constraints its possible instantiations. In this sense,
its position in the list enforces an implicit scoping of the name. 
Bi-traces are essentially observer theories with added structures. The notion of
consistency of bi-traces is therefore based on the notion of consistency for
observer theories, with the added constraint on the possible instantiations
of names in the bi-traces. The latter gives rise to the notion of respectful
substitutions, much like the same notion that appears in the definition of
open bisimulation for the $\pi$-calculus. 

\subsection{Name distinction}
A good definition of open bisimulation for the spi-calculus should naturally
address the issue of name distinction. As in the definition of open bisimulation
for the $\pi$-calculus, the fresh names extruded by a bound output action of a process
should be considered distinct from all other pre-existing names. We employ a syntactic
device to encode this distinction implicitly. We extend the language of processes with
a countably infinite set of {\em rigid names}. Rigid names are basically constants,
so they are not subject to instantiations and therefore cannot be identified by 
substitutions. Note that it is possible to formulate open bisimulation without 
the use of rigid names, at a price of an added complexity.

\paragraph{Outline of the paper} In Section~\ref{sec:spi} we review some notations
and the operational semantics for the spi-calculus. We assume that the reader
has some familiarity with the spi-calculus, so we will not explain in details
the meaning of various constructs of the calculus. 
Section~\ref{sec:obsv} presents the notion of observer theories along with its
various properties. 
Section~\ref{sec:open} defines our notion of open bisimulation, using the bi-trace
structure. A considerable part of this section is devoted to studying properties of bi-traces.
Section~\ref{sec:upto} defines several up-to techniques for open bisimulation.
The main purpose of these techniques is to show that open bisimilarity is closed
under parallel composition, from which we obtain the soundness of open bisimulation
with respect to testing equivalence in Section~\ref{sec:sound}. Section~\ref{sec:ex}
presents some examples of reasoning about bisimulation using the up-to techniques.
Section~\ref{sec:congr} shows that open bisimilarity
is a congruence relation on finite spi-processes without rigid names. 
Section~\ref{sec:conc} concludes the paper and outlines some directions for future work.

\section{The Spi Calculus}
\label{sec:spi}

In this section we review the syntax and the operational semantics
for the spi-calculus. We assume the reader 
has some familiarity with the spi-calculus, so we will not go into details
of the meaning of operators of the spi-calculus. 
We follow the original presentation of the
spi calculus as in \cite{abadi99ic}, but we consider a more restricted
language, i.e., the one with only the pairing and encryption operators.
We assume a denumerable set of names, denoted with ${\cal N}$.
We use $m$, $n$, $x$, $y$, and $z$ to range over names. 
In order to simplify the presentation of open bisimulation, we introduce
another infinite set of names which we call {\em rigid names}, denoted with ${\cal RN}$,
which are assumed to be of a distinct syntactic category from names. 
Rigid names are a purely syntactic device to simplify presentation. It can be
thought of as names which are created when restricted names in processes are
extruded in their transitions. Rigid names embody a notion of {\em distinction},
as in open bisimulation for the $\pi$-calculus~\cite{sangiorgi96acta}, 
in the sense that 
they cannot be instantiated, thus cannot be identified with other rigid names.
The motivation for having rigid names will become clear when we present
open bisimulation in Section~\ref{sec:open}. Rigid names are ranged over by
bold lower-case letters, e.g., as in $\abf$, $\bbf$, $\cbf$, etc. 
We use $u$, $v$, $w$ to range over both names and rigid names.

Messages in the spi calculus are not just names, but can be compound terms,  
for instance encrypted messages. The set of terms is given by the following grammar:
$$
M,N ::= x ~ | ~ \abf \mid \spr M N \mid \enc M N
$$
where $\spr M N$ denotes a pair consisting of messages $M$ and $N$, and
$\enc M N$ denotes the message $M$ encrypted with the key $N$.
The set of processes is defined by the grammar:
$$
\begin{array}{rl}
P,Q,R ::= & 0 \mid \spout M N P \mid M(x).P \mid P | Q \mid (\nu x) P \\
     &  \mid ~ !P ~ \mid [M = N]P \mid \splet x y M P \\
     & \mid \spcase L x N P
\end{array}
$$
The names $x$ and $y$ in the restriction, the `let' and the `case' constructs are binding
occurences. 
We assume the usual $\alpha$-equivalence on process expressions.
The set of terms (messages) is denoted with ${\cal M}$ and the set of processes
with ${\cal P}$.
Given a syntactic expression $E$, e.g., a process, a set of process,
pairs, etc., we write $\freeNm{E}$ to denote the set of free names
in $E$. Likewise, $\freeRN{E}$ denote the set of free rigid names in $E.$
We use the notation $\freeRFN{E}$ to denote
$\freeNm{E} \cup \freeRN{E}.$
We call a process $P$ {\em pure} if there are no free 
occurrences of rigid names in $P.$ The set of pure processes is denoted by
${\cal P}_p.$ Likewise, a message $M$ is pure if $\freeRN{M} = \emptyset$. 
The set of pure messages is denoted by ${\cal M}_p.$

{\em A substitution} is a mapping from names to messages. 
Substitutions are ranged over by $\theta$, $\sigma$ and $\rho$. The domain of substitutions is
defined as $\dom{\theta} = \{x \mid \theta(x) \not = x\}.$ We consider only
substitutions with finite domains. The substitution with empty domain
is denoted by $\epsilon.$ 
We often enumerate the mappings of a substitution
on its finite domain, using the notation $[M_1/x_1, \cdots, M_n/x_n].$
Substitutions are generalised straightforwardly
to mappings between terms (processes, messages, etc.), with the usual proviso
that the free names in the substitutions do not become bound as a result of
the applications of the substitutions. Applications of substitutions to terms 
(processes or messages) are written in postfix notation, e.g., 
as in $M\theta$. Composition of two substitutions $\theta$ and $\sigma$, written $(\theta \circ \sigma)$,
is defined as follows: $M(\theta \circ \sigma) = (M\theta)\sigma.$
Given a substitution $\theta$ and a finite set of  names $V$,
we denote with $\rsubst \theta V$ the substitution 
which coincides with $\theta$ on the set $V$, and is the identity map
everywhere else.

\subsection{Operational semantics}

We use the operational semantics of the spi calculus as it is given
in \cite{abadi98njc}, with one small modification: we allow communication
channels to be arbitrary messages, instead of just names. We do this in order to
get a simpler formulation of open bisimulation in Section~\ref{sec:open}, since
we do not need to keep track of certain constraints related to channel names. 

The one-step transition relations are not relating
processes with processes, rather processes with {\em agents}. The latter is
presented using the notion of {\em abstraction} and {\em concretion} of
processes. Abstractions are expressions of the form
$(x)P$ where $P$ is a process and the construct $(x)$ binds free occurences of
$x$ in $P$, and concretions are expressions of the form
$(\nu \vec x)\langle M \rangle P$ where $M$ is a message and $P$ is a process.
Agents are ranged over by $A$, $B$ and $C.$ As with processes, we call an agent
$A$ {\em pure} if $\freeRN{A} = \emptyset.$

To simplify the presentation of the operational semantics, 
we define compositions between processes and agents as follows.
In the definition below we assume that $x \not \in \{\vec y\} \cup \freeNm{R}$
and $\{\vec y, z\} \cap \freeNm{R} = \emptyset.$
$$
\begin{array}{c}
(\nu x)(z)P \defeq (z) (\nu x)P\\
R ~|~ (x)P \defeq (x)(R~|~P), \hbox{ if $x \not \in \freeNm{R}$}\\
(\nu x)(\nu \vec y)\langle M \rangle Q \defeq (\nu x, \vec y)\langle M\rangle Q,
\hbox{ if $x \in \freeNm{M}$} \\
(\nu x)(\nu \vec y)\langle M \rangle Q \defeq (\nu \vec y)\langle M\rangle (\nu x) Q,
\hbox{ if $x \not \in \freeNm{M}$} \\
R ~|~ (\nu \vec y)\langle M \rangle Q
\defeq (\nu \vec y)\langle M \rangle (R ~ | ~ Q).
\end{array}
$$
The dual composition $A ~|~ R$ is defined symmetrically.

Given an abstraction $F = (x)P$ and a concretion
$(\nu \vec y)\langle M \rangle Q$, where
$\{\vec y\} \cap \freeNm P = \emptyset$,
the {\em interactions} of $F$ and $C$ are defined as follows:
$$
\begin{array}{c}
F @ C \stackrel{\Delta}{=} (\nu \vec y)(P[M/x] ~|~ Q)\\
C @ F \stackrel{\Delta}{=} (\nu \vec y)(Q ~|~ P[M/x]).
\end{array}
$$
We define a reduction relation $>$ on processes as follows:
$$
\begin{array}{rcl}
{!P} & > & P ~|~ {!P} \\
{[M = M] P} & > & P\\
\splet x y {\spr M N} P & > & P[M/x][N/y] \\
\spcase {\enc M N} x N P & > & P[M/x]
\end{array}
$$

\begin{figure}
{\small
$$
\infer[]
{\one {M(x).P} M {(x)P}}
{}
\qquad
\infer[]
{\one{\spout M N P}{\overline M}{\langle N \rangle P}}
{}
$$

$$
\infer[]
{\one{P ~|~ Q}{\tau}{F @ C} }
{\one P M F & \one Q {\over M} C}
\qquad
\infer[]
{\one {P ~|~ Q}{\tau}{C @ F}}
{\one Q {\bar N} C & \one P N F}
$$

$$
\infer[]
{\one{P}{\alpha}{A}}
{P > Q & \one Q \alpha A}
\qquad
\infer[]
{\one {P~|~Q} \alpha {A~|~Q}}
{\one P \alpha A}
$$

$$
\infer[]
{\one {P~|~Q} \alpha {P~|~A}}
{\one Q \alpha A}
\qquad
\infer[]
{\one {(\nu m)P}{\alpha}{(\nu m)A}}
{\one P \alpha A & m \not \in \freeNm{\alpha}}
$$
}
\caption{The operational semantics of the spi calculus.}
\label{fig:spi}
\end{figure}

The operational semantics of the spi calculus is given in Figure~\ref{fig:spi}.
The action $\alpha$ can be either the silent action $\tau$, a term $M$, or a {\em co-term}
$\overline M$, where $M$ is a term. 
We note that as far as the operational semantics is concerned, there is no distinction 
between a name and a rigid name; both can be used as channel names and as messages. 

{\em Structural equivalence} on processes is the least relation 
satisfying the following equations and rules
$$
P ~ | ~ 0 \equiv P,
\quad
P~|~ Q \equiv Q ~ | ~ P,
\quad
P ~ | ~ (Q ~ | ~ R) \equiv (P~|~Q) ~ | ~ R,
$$
$$
(\nu x)(\nu y) P \equiv (\nu y)(\nu x)P,
\quad
(\nu x) 0 \equiv 0, 
\quad
(\nu x)(P ~|~ Q) \equiv P ~|~ (\nu x)Q, \hbox{ if $x \not \in \freeNm{P}$, }
$$
$$
\infer[]
{P \equiv Q}
{P > Q}
\qquad
\infer[]
{P \equiv P}
{}
\qquad
\infer[]
{P \equiv Q}
{Q \equiv P}
$$
$$
\infer[]
{P \equiv R}
{P \equiv Q & Q \equiv R}
\qquad
\infer[]
{P ~|~ Q \equiv P' ~ | ~ Q}
{P \equiv P'}
\qquad
\infer[]
{(\nu m) P \equiv (\nu m) P' }
{P \equiv P'}
$$
Structural equivalence extends to agents by adding the following
rules: 
$$
\infer[]
{(x)P \equiv (x)Q}
{P \equiv Q}
\qquad
\infer[]
{(\nu \vec n) \langle M \rangle P \equiv (\nu \vec m) \langle M\rangle Q}
{P \equiv Q, \hbox{ $\vec m$ is a permutation of $\vec n.$} }
$$

Structurally equivalent processes are indistinguishable as far as their 
transitions are concerned. 
\begin{proposition}
If $P \equiv Q$ then $\one P \alpha {A}$ implies 
$\one Q \alpha {B}$ for some $B$ such that $A \equiv B.$
\end{proposition}
\begin{proof}
By structural induction on the derivations of $P \equiv Q$ and $\one P \alpha A.$
\qed
\end{proof}

\subsection{Testing equivalence}

In order to define testing equivalence, we first define the notion of 
a {\em barb}. A barb is an input or an output channel on which a process can
communicate. We assume that barbs contain no rigid names. 
We denote the reflexive-transitive closure of the silent transition $\stackrel{\tau}{\longrightarrow}$
with  $\stackrel{\tau}{\longrightarrow^*}.$

\begin{definition}
\label{def:testing-equiv}
Two pure processes $P$ and $Q$ are said to be {\em testing equivalent},
written $P \sim Q$, when for every pure process $R$ and every barb $\beta$, if
$$
P ~|~ R \stackrel{\tau}{\longrightarrow^* } P' \stackrel{\beta}{\longrightarrow} A 
$$
for some $P'$ and $A$, then
$$
Q ~|~ R \stackrel{\tau}{\longrightarrow^* } Q' \stackrel{\beta}{\longrightarrow} B 
$$
for some $Q'$ and $B$, and vice versa.
\end{definition}

Notice that testing equivalence is defined for pure processes only,
therefore our definition of testing equivalence coincides with that in  ~\cite{abadi99ic}.





\section{Observer theory}
\label{sec:obsv}

An {\em observer theory} is just a finite set of pairs of
messages, i.e., a subset of ${\cal M} \times {\cal M}.$
The pairs of messages in an observer theory denote the pairs of indistinguishable
messages from the observer point of view. 
An observer theory is essentially what is referred
to as the frame-theory pair in frame bisimulation~\cite{abadi98njc}, i.e.,
the pair $(fr, th)$ where $fr$ is a {\em frame}, i.e., a 
finite set of names and $th$ is a {\em theory}, i.e., a 
finite set of pairs of messages. The frame $fr$ 
represents the names that are known to the observer or environment,
whereas the theory part corresponds to the messages that the observer
obtains through its interaction with a pair of processes. 
Here we adopt the convention that {\em all} names are known
to the observer; rigid names, on the other hand, play the role of  
``private names'', which may or may not be known to the observer.
Thus the ``frame'' component in our observer theory is implicit.

Associated with an observer theory are certain proof systems representing
the deductive capability of the observer. These proof systems allow for
derivation of new knowledge from existing ones. 
Observer theories are ranged over by $\Gamma$ and $\Delta$. We often refer to an observer
theory simply as a {\em theory}. Given a theory $\Gamma$, we write $\pi_1(\Gamma)$ to denote the
set $\{ M \mid \exists N. (M,N) \in \Gamma  \}$, and likewise, $\pi_2(\Gamma)$
to denote the set $\{ N \mid \exists M. (M,N) \in \Gamma\}.$
The observer can encrypt and decrypt messages it has in order to either 
analyze or syntesize messages to deduce the equality of messages. 
This deductive capability is presented as a proof system in Figure~\ref{fig:eqm}. 
This proof system is a straightforward adaptation of the standard proof systems 
for message analysis and synthesis, usually presented in a natural-deduction style, 
e.g., as found in ~\cite{boreale01icalp}, to sequent calculus. 
We find sequent calculus a more natural setting to prove various properties 
of observer theories.
The sequent $\Gamma \seqsym M \eqm N$ means that the messages $M$ and $N$ are indistinguishable
in the theory $\Gamma$. We shall often write $\Gamma \vdash M \eqm N$ to mean 
that the sequent $\Gamma \seqsym M \eqm N$ is derivable using the
rules in Figure~\ref{fig:eqm}. 
Notice that in the proof system in Figure~\ref{fig:eqm}, two names
are indistinguishable if they are syntactically equal. This reflects the fact
that names are entities known to the observer. 

It is useful to consider the set of messages that can be constructed
by an observer in its interaction with a particular process.
This synthesis of messages follows the inference rules given
in Figure~\ref{fig:syn-msg}. The symbol $\Sigma$ denotes a finite set of messages.
We overload the symbols $\seqsym$ and $\vdash$ to denote, respectively, sequents and 
derivability relation of messages given a set of messages.
The rules for message synthesis are just a projection of the rules
for message equivalence.

\begin{lemma}
\label{lm:projected-entailment}
If $\Gamma \vdash M \eqm N$ then $\pi_1(\Gamma) \vdash M$ and
$\pi_2(\Gamma) \vdash N.$
\end{lemma}


\begin{figure}
$$
\infer[var]
{\Gamma \seqsym x \eqm x}
{}
\quad
\infer[id]
{\Gamma, (M,N) \seqsym M \eqm N}
{}
\quad
\infer[pr]
{\Gamma \seqsym \spr M N \eqm \spr {M'} {N'}}
{\Gamma \seqsym M \eqm M' & \quad \Gamma \seqsym N \eqm N'}
$$

$$
\infer[pl]
{\Gamma, (\spr {M_1} {N_1}, \spr{M_2}{N_2}) \seqsym M \eqm N}
{\Gamma, (\spr {M_1} {N_1}, \spr{M_2}{N_2}), 
  (M_1, M_2), (N_1, N_2) \vdash M \eqm N}
$$

$$
\infer[er]
{\Gamma \seqsym \enc M N \eqm \enc {M'} {N'}}
{\Gamma \seqsym M \eqm M' & \quad \Gamma \seqsym N \eqm N'}
$$

$$
\infer[el]
{\Gamma, (\enc {M_1} {N_1}, \enc {M_2} {N_2}) \seqsym M \eqm N}
{\Gamma, (\enc {M_1} {N_1}, \enc {M_2} {N_2}) \seqsym N_1 \eqm N_2 
& \quad 
\Gamma, (\enc {M_1} {N_1}, \enc {M_2} {N_2}), 
(M_1, M_2), (N_1, N_2) \seqsym M \eqm N}
$$

\caption{Proof system for deriving message equivalence}
\label{fig:eqm}
\end{figure}

\begin{figure}
$$
\infer[var]
{\Sigma \seqsym x}
{}
\qquad
\infer[id]
{\Sigma, M \seqsym M}
{}
$$
$$
\infer[pr]
{\Sigma \seqsym \spr M N}
{\Sigma \seqsym M & \Sigma \seqsym N}
\qquad
\infer[er]
{\Sigma \seqsym \enc M N}
{\Sigma \seqsym M & \Sigma \seqsym N}
$$
$$
\infer[pl]
{\Sigma, \spr M N \seqsym R}
{\Sigma, \spr M N, M, N \seqsym R}
\qquad
\infer[el]
{\Sigma, \enc M N \seqsym R}
{\Sigma, \enc M N \seqsym N & \Sigma, \enc M N, M, N \seqsym R}
$$
\caption{Proof system for message synthesis}
\label{fig:syn-msg}
\end{figure}

A nice feature of the sequent calculus formulation is 
that it satisfies the so-called ``sub-formula property'', that is, 
in any derivation of a judgment, every judgment in the derivation
contains only subterms occuring in the judgment at the root of the derivation tree.
This gives us immediately a bound on the depth of the derivation tree, hence
the decidability of the proof systems. 

\begin{proposition}
Given any $\Gamma$, $\Sigma$, $M$ and $N$, it is decidable whether the judgments
$\Gamma \vdash M \eqm N$ and $\Sigma \vdash M$ hold. 
\end{proposition}

\subsection{Properties of the entailment relations}

We examine several general properties of the entailment relation
$\vdash$ which will be used throughout the paper.

The following two lemmas state that the rules for $\eqm$ are invertible,
under some conditions. Lemma~\ref{lm:invertible} actually states something
stronger than just invertibility; it also says that keeping the components
of a message pair instead of the compound pair amounts to the same thing,
again under a certain condition. This stronger statement, if coupled
with the weakening lemma (Lemma~\ref{lm:weakening}), trivially entails
the invertibility of left-rules under the given condition.
The proofs of the next two lemmas are straightforward by induction
on the length of derivations.

\begin{lemma}
\label{lm:invertible}
The sequent
$$\Gamma, (\spr {M_1} {N_1}, \spr {M_2} {N_2}) \seqsym M \eqm N$$
is derivable if and only if 
$$\Gamma, (M_1, M_2), (N_1, N_2) \seqsym M \eqm N$$ is derivable.
If $\Gamma, (\enc {M_1} {N_1}, \enc {M_2} {N_2}) \vdash N_1 \eqm N_2$, 
then
$$\Gamma, (\enc {M_1} {N_1}, \enc {M_2} {N_2}) \seqsym M \eqm N$$ 
is derivable if and only if 
$$\Gamma, (M_1, M_2), (N_1, N_2) \seqsym M \eqm N$$ is derivable.
\end{lemma}

\begin{lemma}
\label{lm:right-invertible}
The judgment $\Gamma \seqsym \spr {R} {T} \eqm \spr U V$ is 
derivable if and only if $\Gamma \seqsym R \eqm U$ and
$\Gamma \seqsym T \eqm V$ are derivable.
If $\Gamma \vdash T \eqm V$ then 
$\Gamma \seqsym \enc R T \eqm \enc U V$ is derivable if and only if
$\Gamma \seqsym R \eqm U$ is derivable. 
\end{lemma}

The next two lemmas show that the entailment relation $\vdash$ for 
message equivalence and synthesis are monotonic.

\begin{lemma}
\label{lm:weakening}
If $\Gamma \vdash M \eqm N$ then 
$\Gamma, (R,T) \vdash M \eqm N$ for any $(R,T).$
If $\Sigma \vdash M$ then 
$\Sigma, R \vdash M$ for any $R$.
\end{lemma}

\begin{lemma}
\label{lm:name-weakening}
$\Gamma \vdash M \eqm N$ if and only if 
$(x,x), \Gamma \vdash M \eqm N$, for any $\Gamma$, $M$, $N$ and $x.$
\end{lemma}

\begin{lemma}
\label{lm:inverse}
If $\Gamma \vdash M \eqm N$ then $\Gamma^{-1} \vdash N \eqm M.$
\end{lemma}

The following proposition states the transitivity of the entailment relation.
Readers familiar with proof theory will recognize its similarity to 
the ``cut-elimination'' theorem. 

\begin{proposition}
\label{prop:cut-elimination}
If $\Gamma \vdash M \eqm N$ and $\Delta, (M,N) \vdash R \eqm T$
then $\Gamma \cup \Delta \vdash R \eqm T.$
\end{proposition}
\begin{proof}
Suppose $\Pi_1$ is the derivation of $\Gamma \seqsym M \eqm N$
and $\Pi_2$ is the derivation of $\Delta, (M,N) \seqsym R \eqm T.$
We show that there exists
a derivation $\Pi$ of $\Gamma \cup \Delta \seqsym R \eqm T.$
The proof is by induction on the height of $\Pi_1.$ 
We distinguish several cases based on the last rules in $\Pi_1$. 
We first note that if $(M,N) \in \Delta$ then $\Pi$ can be constructed
directly from $\Pi_2$ by applying the weakening lemma 
(Lemma~\ref{lm:weakening}). In the following we assume that 
$(M,N) \not \in \Delta.$
\begin{enumerate}
\item $\Pi_1$ ends with the $var$-rule. In this case, $\Pi_2$ is
a derivation of $(x,x), \Delta \seqsym R \eqm T.$
Hence, by Lemma~\ref{lm:weakening} and Lemma~\ref{lm:name-weakening}, we have 
$\Gamma \cup \Delta \vdash R \eqm T$ as well.

\item $\Pi_1$ ends with the $id$-rule. In this case, 
$(M,N) \in \Gamma$, hence $(M,N) \in \Gamma \cup \Delta$.
Applying Lemma~\ref{lm:weakening} to $\Pi_2$, we obtain a derivation
of $\Gamma \cup \Delta \seqsym R \eqm T$ as required.

\item $\Pi_1$ ends with $pl$:
$$
\infer[pl]
{\Gamma', (\spr U V, \spr X Y) \seqsym M \eqm N}
{\deduce{\Gamma', (U, X), (V, Y) \seqsym M \eqm N}{\Pi_1'}}
$$
By the induction hypothesis, we have a derivation $\Pi'$ of
$$
\{\Gamma', (U,X), (V,Y) \} \cup \Delta \seqsym R \eqm T.
$$
The derivation $\Pi$ is therefore obtained from $\Pi'$ 
by applying the $pl$-rule to the pairs $(U,X)$ and $(V, Y).$

\item $\Pi_1$ ends with $el$:
$$
\infer[el]
{\Gamma', (\enc U V, \enc X Y) \seqsym M \eqm N}
{\deduce{\Gamma \seqsym V \eqm Y}{\Pi_3} & 
\deduce{\Gamma, (U, X), (V, Y) \seqsym M \eqm N}{\Pi_4}}
$$
By the induction hypothesis (on $\Pi_4$) we have a derivation $\Pi'$
of $\{\Gamma, (U,X), (V,Y)\} \cup \Delta \seqsym R \eqm T,$
and applying Lemma~\ref{lm:weakening} to $\Pi_3$ we obtain a derivation
$\Pi_3'$ of $\Gamma \cup \Delta \seqsym V \eqm Y.$
The derivation $\Pi$ is then constructed as follows:
$$
\infer[el]
{\Gamma \cup \Delta \seqsym R \eqm T}
{
\deduce{\Gamma \cup \Delta \seqsym V \eqm Y}{\Pi_3'}
&
\deduce{\{\Gamma, (U,X), (V,Y)\} \cup \Delta \seqsym R \eqm T}{\Pi'}
}
$$

\item $\Pi_1$ ends with the $pr$-rule:
$$
\infer[pr]
{\Gamma \seqsym \spr {M_1}{M_2} \eqm \spr {N_1}{N_2}}
{
\deduce{\Gamma \seqsym M_1 \eqm N_1}{\Pi_1'}
&
\deduce{\Gamma \seqsym M_2 \eqm N_2}{\Pi_1''}
}
$$
Applying Lemma~\ref{lm:invertible} to $\Pi_2$, we obtain a derivation
$\Pi_2'$ of 
$$
\Delta, (M_1, M_2), (N_1, N_2) \seqsym R \eqm T.
$$
The derivation $\Pi$ is then constructed by applying the induction hypothesis
twice (one on $\Pi_1'$ and the other on $\Pi_1''$).

\item $\Pi_1$ ends with the $er$-rule:
$$
\infer[pr]
{\Gamma \seqsym \enc {M_1}{M_2} \eqm \enc {N_1}{N_2}}
{
\deduce{\Gamma \seqsym M_1 \eqm N_1}{\Pi_1'}
&
\deduce{\Gamma \seqsym M_2 \eqm N_2}{\Pi_1''}
}
$$
Applying Lemma~\ref{lm:weakening} to $\Pi_1''$ and $\Pi_2$, we obtain
two derivations:
$$
\deduce{\Gamma \cup \Delta, (\enc {M_1}{M_2}, \enc {N_1}{N_2}) 
\seqsym M_2 \eqm N_2}{\Pi_3}
\qquad
\hbox{ and }
\qquad
\deduce{\Gamma \cup \Delta, (\enc {M_1}{M_2}, \enc {N_1}{N_2}) 
\seqsym R \eqm T.}{\Pi_4}
$$
Therefore, by Lemma~\ref{lm:invertible}, we have a derivation,
say $\Pi'$ of 
$$
\Gamma \cup \Delta, (M_1,N_1), (M_2, N_2) \seqsym R \eqm T.
$$
The derivation $\Pi$ is then constructed by applying the induction
hypothesis twice, that is, by first cutting $\Pi_1'$ with
$\Pi'$, followed by another cut with $\Pi_1''$. 
\end{enumerate}
\qed
\end{proof}

\subsection{Consistency of observer theory}

Recall that the motivation behind the notion of message equivalence $\eqm$
is for it to replace syntactic equality in the definition of bisimulation.
This would require that the relation $\eqm$ to satisfy certain properties,
e.g., a uniqueness property like $M \eqm N$ and $M \eqm N'$
implies $N = N'.$ Since the relation $\eqm$ is parameterised upon an observer theory,
we shall investigate under what conditions an observer theory gives rise to a well-behaved
relation $\eqm.$ In the literature of bisimulation for spi calculus,
this notion is usually referred to as the {\em consistency} property of 
observer theories (or other structures encoding the environment's knowledge).
We now define an abstract notion of theory consistency, based on
the entailment relation $\vdash$ defined previously. 
We later show that this abstract notion of consistency is equivalent
to a more concrete one which is finitely checkable.

\begin{definition}
\label{def:theory-consistency}
A theory $\Gamma$ is {\em consistent} if for every $M$ and $N$, if 
$\Gamma \vdash M \eqm N$ then the following hold:
\begin{enumerate}
\item $M$ and $N$ are of the same type of expressions, i.e., $M$ is
a pair (an encrypted message, a (rigid) name) if and only if $N$ is.
\item If $M = \enc {M_1} {M_2}$ and $N = \enc {N_1}{N_2}$ then
$\pi_1(\Gamma) \vdash M_2$ implies $\Gamma \vdash M_2 \eqm N_2$ and
$\pi_2(\Gamma) \vdash N_2$ implies $\Gamma \vdash M_2 \eqm N_2$.
\item For any $R$, $\Gamma \vdash M \eqm R$ implies $R = N$ 
and $\Gamma \vdash R \eqm N$ implies $R = M.$
\end{enumerate}

\end{definition}

The first condition in Definition~\ref{def:theory-consistency} states that
the equality relation $\eqm$ respects types, i.e., it is not possible
that an operation (pairing, encryption) on $M$ succeeds while the same
operation on $N$ fails. The second condition states
that both projections of the theory contain ``equal'' amount of knowledge, e.g.,
it is not possible that one message decrypts while the other fails to. 
The third condition states the unicity of $\eqm.$
Note that consistent theories always entail $x \eqm x$ for any name $x$. 


\subsection{A finite characterisation of consistent theories}


The notion of consistency as defined in Definition~\ref{def:theory-consistency}
is not obvious to check since it involves quantification over all 
equivalent pairs of messages. We show that a theory can be reduced
to a certain normal form for which there exist finitely checkable
properties that entail consistency of the original theory.
For this purpose, we define a rewrite relation
on theories. 

\begin{definition}
\label{def:rewrite}
The rewrite relation $\longrightarrow$ on observer theories is defined as follows:
$$
\begin{array}{rcl}
\Gamma, (\spr M N, \spr {M'} {N'}) & \longrightarrow & \Gamma, (M, M'), (N, N') \\
\Gamma, (\enc M N, \enc {M'} {N'}) & \longrightarrow & \Gamma, (M, M'), (N, N') \\
   & & \mbox{ if $\Gamma, (\enc M N, \enc {M'} {N'})  \vdash N \eqm N'$.} \\
\end{array}
$$
A theory $\Gamma$ is {\em irreducible} if $\Gamma$ cannot be rewritten
to any other theory. 
$\Gamma$ is an irreducible form of another theory $\Gamma'$ if $\Gamma$ is 
irreducible and $\Gamma' \longrightarrow^* \Gamma$. 
\end{definition}

\begin{lemma}
\label{lm:weakening-consistency}
If $\Gamma$ is consistent and $\Gamma \vdash M \eqm N$ then
$\Gamma \cup \{(M,N) \}$ is consistent.
\end{lemma}

\begin{lemma}
Every observer theory $\Gamma$ has a unique irreducible form.
\end{lemma}
\begin{proof}
Since the rewrite system is obviously terminating, it is enough 
to show that it is locally confluent, that is, 
if $\Gamma \longrightarrow \Gamma_1$ and $\Gamma \longrightarrow
\Gamma_2$ then there exists $\Gamma_3$ such that 
$\Gamma_1 \longrightarrow^* \Gamma_3$ and
$\Gamma_2 \longrightarrow^* \Gamma_3.$
There are no critical pairs in the rewrite system. We need only to
verify that the side condition of the rewrite rules is not affected by
the different sequences of rewrites, which is a simple corollary of
Lemma~\ref{lm:invertible}.
We show here one case involving encryption, the other cases
are straightforward. 
Suppose we have two possible rewrites:
$$
\Gamma = \Gamma',  (\enc {R_1}{T_1}, \enc {R_2}{T_2}), 
(\enc {M_1} {N_1}, \enc {M_2}{N_2}) \longrightarrow 
\Gamma', (\enc {R_1}{T_1}, \enc {R_2}{T_2}), (M_1, M_2), (N_1, N_2) = \Gamma_1
$$
where $\Gamma  \vdash N_1 \eqm N_2$, and 
$$
\Gamma', (\enc {R_1}{T_1}, \enc {R_2}{T_2}), (\enc {M_1} {N_1}, \enc {M_2}{N_2}) \longrightarrow
\Gamma', (R_1, R_2), (T_1, T_2), (\enc {M_1} {N_1}, \enc {M_2}{N_2}) = \Gamma_2,
$$
where $\Gamma \vdash {T_1}  \eqm {T_2}$.
Let $\Gamma_3$ be the theory $\Gamma', (R_1,R_2), (T_1,T_2), (M_1, M_2), (N_1,N_2)$.
By Lemma~\ref{lm:invertible}, we have $\Gamma_1 \vdash T_1 \eqm T_2$
and $\Gamma_2 \vdash N_1 \eqm N_2$, and therefore
$$
\Gamma_1 \longrightarrow \Gamma_3 \longleftarrow \Gamma_2.
$$
\qed
\end{proof}

We denote the irreducible form of $\Gamma$ with $\irred{\Gamma}$.
The irreducible form is equivalent to $\Gamma$, in the sense that
they entail the same set of equality of messages.

\begin{lemma}
\label{lm:invertible2}
If $\Gamma \longrightarrow \Gamma'$ 
then $\Gamma \vdash M \eqm N$ if and only if 
$\Gamma' \vdash M \eqm N.$
\end{lemma}
\begin{proof}
This is a simple corollary of Lemma~\ref{lm:invertible}.
\qed
\end{proof}

The reduction on observer theories also preserves the set of messages entailed
by their projections. 

\begin{lemma}
\label{lm:projected-reduct}
Suppose $\Gamma \longrightarrow \Gamma'$. Then for all $M$,
$\pi_i(\Gamma) \vdash M$ if and only if $\pi_i(\Gamma') \vdash M$.
\end{lemma}
\begin{proof}
Straightforward from the definition of reduction on theories and
simple induction on the length of proofs on the entailment relation.
\qed
\end{proof}

An immediate consequence of the above lemma is the following.
\begin{lemma}
For all $M$ and for all $\Gamma$, $\pi_i(\Gamma) \vdash M$ if and only if
$\pi_i(\irred{\Gamma}) \vdash M$.
\end{lemma}

\begin{lemma}
\label{lm:rewrite-consistent}
If $\Gamma \longrightarrow^* \Gamma'$,
then $\Gamma$ is a consistent if and only if $\Gamma'$ is consistent.
\end{lemma}
\begin{proof}
By Lemma~\ref{lm:invertible2} and Lemma~\ref{lm:projected-reduct}, 
the rewrite rule preserves derivability
of equations and synthesis of messages in both ways. Therefore 
the properties of consistency in Definition~\ref{def:theory-consistency}
are preserved by the reduction.
\qed
\end{proof}

\begin{lemma}
\label{lm:reduced-consistent}
A theory $\Gamma$ is consistent if and only if $\irred{\Gamma}$ is
consistent.
\end{lemma}
\begin{proof}
This is a simple corollary of Lemma~\ref{lm:rewrite-consistent}. \qed
\end{proof}

We are now ready to state the finite characterisation of consistent theories.

\begin{proposition}
\label{prop:characterisation-of-consistency}
A theory $\Gamma$ is consistent if and only if $\irred{\Gamma}$ satisfies
the following conditions: if $(M,N) \in \irred{\Gamma}$ then
\begin{description}
\item[(a)] $M$ and $N$ are of the same type of expressions, in particular,
if $M = x$, for some name $x$, then $N = x$ and vice versa, 
\item[(b)] if $M = \enc {M_1} {M_2}$ and $N = \enc {N_1}{N_2}$ then
$\pi_1(\irred \Gamma) \not \vdash M_2$ and  
$\pi_2(\irred \Gamma) \not \vdash N_2$.
\item[(c)] for any $(U,V) \in \irred{\Gamma}$, $U = M$ if and only if $V = N$.
\end{description}
\end{proposition}
\begin{proof}
Suppose that $\Gamma$ is consistent. We show that $\irred \Gamma$ 
satisfies (a), (b) and (c). By Lemma~\ref{lm:reduced-consistent}, 
$\irred \Gamma$ is consistent. The criteria (a) and (c) 
follows straightforwardly from Definition~\ref{def:theory-consistency} (1) and (3). 
To show (b), suppose that $M = \enc{M_1}{M_2}$ and $N = \enc {N_1} {N_2}$ but
$\pi_1(\irred \Gamma) \vdash M_2$. 
By Definition~\ref{def:theory-consistency}(2), we have $\irred \Gamma \vdash M_2 \eqm N_2.$
But this entails that $\irred \Gamma$ is reducible, contrary to the fact that
$\irred \Gamma$ is irreducible. Therefore it must be the case that 
$\pi_1(\irred \Gamma) \not \vdash M_2$. Using a similar argument we can show that
$\pi_2(\irred \Gamma) \not \vdash N_2.$

Now suppose that $\irred \Gamma$ satisfies (a), (b) and (c). We show that
$\Gamma$ is consistent. By Lemma~\ref{lm:reduced-consistent}, 
it is enough to show that
$\irred \Gamma$ is consistent. That is, we show that whenever 
$\irred \Gamma \vdash M \eqm N$, $M$ and $N$ satisfy the conditions
(1), (2) and (3) in Definition~\ref{def:theory-consistency}. This is proved
by induction on the length of the deduction of 
$\irred \Gamma \vdash M \eqm N$. Note that since $\irred \Gamma$ is
irreducible, the derivation $\irred \Gamma \vdash M \eqm N$ does not make
any use of left-rules. 
\begin{enumerate}
\item {\em $M$ and $N$ are of the same type of expressions.} 
This fact is easily
shown by induction on the length of proofs of $\irred \Gamma \vdash M \eqm N.$

\item {\em If $M = \enc {M_1} {M_2}$ and $N = \enc {N_1} {N_2}$ then
$\pi_1(\irred \Gamma) \vdash M_2$ implies $\irred \Gamma \vdash M_2 \eqm N_2$
and 
$\pi_2(\irred \Gamma) \vdash N_2$ implies $\irred \Gamma \vdash M_2 \eqm N_2$.}
We show here a proof of the first part of the conjunction; the other part is
symmetric. The proof is by induction on the length of derivation of 
$\irred \Gamma ~ \vdash M \eqm N.$
Note that since left-rules are not applicable, there are only two possible 
cases to consider.
The first is that $(M, N) \in \irred \Gamma$. In this case, $\pi_1(\irred \Gamma) \not \vdash M_2$,
by the assumption (b) of the statement of the lemma, so the property holds vacuously.
The other case is when the last rule of 
$\irred \Gamma \vdash M \eqm N$ is an encryption rule:
$$
\infer[er]
{\irred \Gamma \seqsym \enc {M_1}{M_2} \eqm \enc {N_1}{N_2}}
{\irred \Gamma \seqsym M_1 \eqm N_1 & \irred \Gamma \seqsym M_2 \eqm N_2}
$$
The property holds trivially, since $\irred \Gamma \vdash M_2 \eqm N_2$.

\item {\em For any $R$, $\irred \Gamma \vdash M \eqm R$ implies
$R = N$ and $\irred \Gamma \vdash R \eqm N$ implies $R = M$.}
We show only the first part of the conjunction; the other part is 
symmetric. We first note that by property (1) above, $M$, $R$ and $N$
must all be of the same type of expressions. The proof is by induction
on the size of $R$:
\begin{itemize}
\item $R = x$, for some name $x$. Then obviously $M = N = R = x$.
\item $R = \abf$, for some rigid name $\abf$. In this case, it
must be the case that $(M, R) \in \irred \Gamma$ and 
$(M,N) \in \irred \Gamma$. Therefore, by the condition (c) in the statement
of the lemma, we have $R = N$.
\item $R = \spr {R_1} {R_2}$. In this case, $M$ and $N$ must also be
pairs, say, $\spr {M_1} {M_2}$ and $\spr {N_1} {N_2}$, and
the derivations of $\irred \Gamma \vdash M \eqm R$ and 
$\irred \Gamma \vdash M \eqm N$ must end with instances of the $pr$-rule.
Therefore we have $\irred \Gamma \vdash R_1 \eqm M_1$,
$\irred \Gamma \vdash R_2 \eqm M_2$, $\irred \Gamma \vdash M_1 \eqm N_1$
and $\irred \Gamma \vdash M_2 \eqm N_2$. By induction hypothesis,
we have $R_1 = N_1$ and $R_2 = N_2$, therefore $R = N$.

\item $R =\enc {R_1} {R_2}$. In this case we have that $M = \enc {M_1} {M_2}$
and $N = \enc {N_1}{N_2}$ for some $M_1$, $M_2$, $N_1$ and $N_2$.
There are two cases to consider here. The first is when 
the derivation of $\irred \Gamma \vdash M \eqm R$ ends with the $id$-rule,
that is, $(M,R) \in \irred \Gamma$. In this case, we argue that 
$(M,N)$ must also be in $\irred \Gamma$: Suppose this is not the case, 
then $\irred \Gamma \vdash M \eqm N$ must end with the $er$-rule, and
as a consequence, $\irred \Gamma \vdash M_2 \eqm N_2$ and 
$\pi_1(\irred \Gamma) \vdash M_2.$ 
By the property (2) above, this entails $\irred  \Gamma \vdash M_2 \eqm R_2$.
But this would mean that $\irred \Gamma$ is reducible, contrary to the 
the fact that $\irred \Gamma$ is irreducible. Hence $(M,N)$ must also be 
in $\irred \Gamma$. Now by the condition (c) in the assumption of the lemma,
we have $R = N$.

The second case is when $\irred \Gamma \vdash M \eqm R$ ends with the
$er$-rule. This case is proved straightforwardly by induction hypothesis.
\end{itemize}
\end{enumerate}
\qed
\end{proof}

Finally, we show that the inverse operation on an observer theory preserves consistency.

\begin{lemma}
\label{lm:inverse-consistency}
If $\Gamma$ is consistent then $\Gamma^{-1}$ is also consistent.
\end{lemma}
\begin{proof}
This follows from Lemma~\ref{lm:inverse} and the definition of
consistency.
\qed
\end{proof}

\subsection{Closure under substitutions}
In the definition of open bisimulation in Section~\ref{sec:open}, we shall consider
substitutions of free names in processes and theories. 
It is crucial that open bisimulation is closed under certain substitutions
in order to show that it is a congruence.
A key technical lemma to prove this congruence property is that
derivability of messages equivalence must be closed under 
a certain class of substitutions. 

The entailment relation $\vdash$ is in general not closed under
arbitrary substitutions, the reason being the inclusion of the 
rule
$$
\infer[var]
{\Gamma \seqsym x \eqm x}
{}
$$
Using this rule, we can prove, for instance, $\emptyset \vdash x \eqm x$.
Now if we substitute $\abf$ for $x$, where $\abf$ is some rigid name,
we do not have $\emptyset \vdash \abf \eqm \abf$, since the $var$-rule does not
apply to rigid names.

We first study a subset of $\vdash$ without the $var$-rule, which we
call $\vdash_c$ (for ``closed'' entailment relation), and show how
this can be used to characterize the kind of substitutions required
for proving closure under substitutions for the entailment
relation $\vdash$.
We shall often work with substitution pairs in the following sections.
Application of a substitution pair $\vec \theta = (\theta_1,\theta_2)$
to a pair of terms $(M,N)$ is defined to be $(M\theta_1,N\theta_2)$.
This extends straightforwardly to application of substitution pairs
to sets or lists of pairs.

The proofs for the following two lemmas are straightforward by induction
on the length of derivations.

\begin{lemma}
\label{lm:names-ext}
Let $\Gamma \vdash M \eqm N$ and let $x_1,\ldots,x_n$ be the free names
in $\Gamma$, $M$ and $N$. Then we have
$$
(x_1,x_1), \ldots, (x_n, x_n), \Gamma \vdash_c M \eqm N.
$$
\end{lemma}

\begin{lemma}
\label{lm:clo-subst1}
If $\Gamma \vdash_c M \eqm N$ then for any substitution pair 
$\vec \theta = (\theta_1,\theta_2)$, 
$\Gamma \vec\theta \vdash_c M\theta_1 \eqm N\theta_2$.
\end{lemma}

\begin{lemma}
\label{lm:clo-subst2}
Let $\Gamma \vdash M \eqm N$ and let $\vec \theta = (\theta_1,\theta_2)$
be a substitution pair such that for all $x \in \freeFN{\Gamma,M,N}$
it holds that
$\Gamma\vec\theta \vdash \theta_1(x) \eqm \theta_2(x).$
Then $\Gamma\vec\theta \vdash M\theta_1 \eqm N\theta_2$.
\end{lemma}
\begin{proof}
Suppose $\freeFN{\Gamma,M,N} = \{x_1,\cdots,x_n\}.$
From Lemma~\ref{lm:names-ext}, we have
$$
(x_1,x_1),\ldots,(x_n,x_n),\Gamma \vdash_c M \eqm N,
$$
and applying Lemma~\ref{lm:clo-subst1} we get
$$
(x_1\theta_1,x_1\theta_2), \ldots, (x_n\theta_1, x_n\theta_2),
\Gamma\vec\theta \vdash_c M\theta_1 \eqm N\theta_2.
$$
Since $\vdash_c ~ \subseteq ~ \vdash$, we also have
$$
(\theta_1(x_1),\theta_2(x_1)), \ldots, (\theta_1(x_n), \theta_2(x_n)),
\Gamma\vec\theta \vdash M\theta_1 \eqm N\theta_2.
$$
From the assumption, we have $\Gamma\vec\theta \vdash \theta_1(x_i) \eqm
\theta_2(x_i)$, for any $i \in \{1,\dots,n\}$.
Therefore, applying Proposition~\ref{prop:cut-elimination} $n$-times, we obtain
$$
\Gamma\vec\theta \vdash M\theta_1 \eqm N\theta_2.
$$
\qed
\end{proof}

\subsection{Composition of observer theories}

\begin{definition}
Let $\Gamma_1$ and $\Gamma_2$ be observer theories. 
$\Gamma_1$ is {\em left-composable} with $\Gamma_2$, or equivalently,
$\Gamma_2$ is {\em right-composable} with $\Gamma_1$, 
if they are of the form
$$
\Gamma_1 = \{(M_1, N_1), \cdots, (M_k, N_k) \}
$$
$$
\Gamma_2 = \{(N_1, R_1), \cdots, (N_k, R_k) \}
$$
and $N_1,\dots, N_k$ are pairwise distinct messages.
Their (unique) composition, denoted by $\Gamma_1 \circ \Gamma_2$, is the theory
$$
\{(M_1, R_1), \cdots, (M_k, R_k) \}.
$$
\end{definition}

\begin{lemma}
\label{lm:trans-eq}
Let $\Gamma_1$ and $\Gamma_2$ be consistent observer theories such that $\Gamma_1$ is left-composable
with $\Gamma_2$. If $\Gamma_1 \vdash M \eqm R$ and $\Gamma_2 \vdash R \eqm N$
then $\Gamma_1 \circ \Gamma_2 \vdash M \eqm N.$
\end{lemma}
\begin{proof}
We prove this by induction on the length of the derivation of $\Gamma_1 \vdash M \eqm R.$

{\em Base cases:} If $M = x$ then $R = x$ and $N = x$, and trivially
$\Gamma_1\circ \Gamma_2 \vdash x \eqm x.$
Otherwise $(M, R) \in \Gamma_1$. Since $\Gamma_1$ and $\Gamma_2$ are composable,
there is a unique $T$ such that $(R, T) \in \Gamma_2$. 
By Definition~\ref{def:theory-consistency}(3), this means that $T = N$. Therefore
we have $(M, N) \in \Gamma_1 \circ \Gamma_2$, hence 
$\Gamma_1 \circ \Gamma_2 \vdash M \eqm N.$

{\em Inductive cases:}
We distinguish several cases based on the last rule in the derivation
of $\Gamma_1 \vdash M \eqm R$. We show here only the cases involving encryptions;
the other cases follow straightforwardly from induction hypothesis.

\begin{itemize}
\item Suppose the last rule is $el$: 
$$
\infer[el]
{\Gamma_1', (\enc S T, \enc U V) \seqsym  M \eqm R}
{
\Gamma_1 \seqsym T \eqm V & \Gamma_1, (S,U), (T,V) \seqsym M \eqm R
}
$$
In this case there must be a pair $(\enc U V, \enc X Y)$ in $\Gamma_2$.
Since $\Gamma_1 \vdash T \eqm V$ and $\pi_1(\Gamma_2) = \pi_2(\Gamma_1)$,
we have that $\pi_1(\Gamma_2) \vdash V$, and by Definition~\ref{def:theory-consistency}(2),
$\Gamma_2 \vdash V \eqm Y$, and by induction hypothesis we have
$$
\Gamma_1 \circ \Gamma_2 \vdash T \eqm Y.
$$
Since $\Gamma_2 \vdash R \eqm N$ and $\Gamma_2 \vdash V \eqm Y$, 
by Lemma~\ref{lm:invertible} and Lemma~\ref{lm:weakening},
we have $\Gamma_2, (U, X), (V, Y) \vdash R \eqm N.$
Since $\Gamma_2$ is consistent and $\Gamma_2 \vdash U \eqm X$ and $\Gamma_2 \vdash V \eqm Y$,
by Lemma~\ref{lm:weakening-consistency} $\Gamma_2 \cup \{(U,X), (V, Y)\}$ is also
consistent. By a similar argument, we can show that
$\Gamma_1 \cup \{(S,U), (T,V) \}$ is consistent.
We can therefore apply the induction hypothesis to get the derivation
$$
\Gamma_1 \circ \Gamma_2, (S, X), (T, Y) \vdash M \eqm N.
$$
The sequent $\Gamma_1 \circ \Gamma_2 \vdash M \eqm N$ can therefore be derived as follows:
$$
\infer[el]
{\Gamma_1 \circ \Gamma_2 \seqsym M \eqm N}
{\Gamma_1 \circ \Gamma_2 \seqsym T \eqm Y 
& \Gamma_1\circ \Gamma_2, (S, X), (T,Y) \seqsym M \eqm N}
$$
where the derivations for the premise sequents are constructed as discussed above.

\item Suppose the last rule is $er$:
$$
\infer[er]
{\Gamma_1 \seqsym \enc {M_1}{M_2} \eqm \enc {R_1}{R_2}}
{
\Gamma_1 \seqsym M_1 \eqm R_1 & \Gamma_1 \seqsym M_2 \eqm R_2
}
$$
Since $\Gamma_2$ is consistent, it must be the case that $N = \enc {N_1}{N_2}$
for some $N_1, N_2.$
Since $\pi_1(\Gamma_2) = \pi_2(\Gamma_1)$, we have $\pi_1(\Gamma_2) \vdash R_2$,
therefore by Definition~\ref{def:theory-consistency}(2), $\Gamma_2 \vdash R_2 \eqm N_2.$
It follows from Lemma~\ref{lm:right-invertible} that 
$\Gamma_2 \vdash R_1 \eqm N_1$ as well. We can therefore apply the induction hypothesis
to obtain
$$
\Gamma_1 \circ \Gamma_2 \vdash M_1 \eqm N_1 \qquad \hbox{ and } \qquad
\Gamma_1 \circ \Gamma_2 \vdash M_2 \eqm N_2,
$$
from which we derive $\Gamma_1 \circ \Gamma_2 \vdash M \eqm N$ by an application of
the $er$-rule.
\end{itemize}
\qed
\end{proof}

\begin{lemma}
\label{lm:reduced-comp1}
Let $\Gamma_1$ and $\Gamma_2$ be consistent theories such that $\Gamma_1$ is left-composable
with $\Gamma_2$. If $\Gamma_1 \circ \Gamma_2 \longrightarrow \Gamma'$ then
there exists $\Gamma_1'$ and $\Gamma_2'$ such that $\Gamma_1'$ is left-composable with
$\Gamma_2'$, 
$\Gamma_1 \longrightarrow \Gamma_1'$, $\Gamma_2 \longrightarrow \Gamma_2'$ and
$\Gamma' = \Gamma_1' \circ \Gamma_2'$.
\end{lemma}
\begin{proof}
We prove this by case analysis on the rewrite step 
$\Gamma_1 \circ \Gamma_2 \longrightarrow \Gamma'.$
The case where the rewrite happens on paired-messages is trivial. We consider the
more difficult case with encryption. 
Suppose $\Gamma_1 = \Gamma_3 \cup \{(\enc R T, \enc U V) \}$
and $\Gamma_2 = \Gamma_4 \cup \{ (\enc U V, \enc M N) \},$ and suppose the rewrite
step is 
$$
\Gamma_1 \circ \Gamma_2 = \Gamma_3 \circ \Gamma_4, (\enc R T, \enc M N) \longrightarrow
\Gamma_3 \circ \Gamma_4, (R, M), (T, N) = \Gamma'
$$ 
where $\Gamma_1 \circ \Gamma_2 \vdash T \eqm N.$
Since $\pi_1(\Gamma_1) = \pi_1(\Gamma_1\circ \Gamma_2)$ and
$\pi_2(\Gamma_2) = \pi_2(\Gamma_1 \circ \Gamma_2)$, we have 
$$
\pi_1(\Gamma_1) \vdash T \qquad
\hbox{ and } \qquad
\pi_2(\Gamma_2) \vdash N.
$$
Since $\Gamma_1$ and $\Gamma_2$ are consistent, by Definition~\ref{def:theory-consistency}(2),
together with the above two facts, we have 
$$
\Gamma_1 \vdash T \eqm V \qquad
\hbox{ and } \qquad
\Gamma_2 \vdash V \eqm N.
$$
Therefore,
$$
\Gamma_1 \longrightarrow \Gamma_3, (R, U), (T, V) = \Gamma_1'
\qquad \hbox{ and } \qquad
\Gamma_2 \longrightarrow \Gamma_4, (U, M), (V, N) = \Gamma_2'.
$$
Obviously, $\Gamma' = \Gamma_1' \circ \Gamma_2'$.
\qed
\end{proof}

\begin{lemma}
\label{lm:reduced-comp2}
Let $\Gamma_1$ and $\Gamma_2$ be consistent theories such that $\Gamma_1$ is left-composable
with $\Gamma_2$. If $\Gamma_1 \circ \Gamma_2$ is irreducible then so are 
$\Gamma_1$ and $\Gamma_2$.
\end{lemma}
\begin{proof}
Suppose $\Gamma_1 \circ \Gamma_2$ is irreducible but $\Gamma_1$ is reducible.
We first show that in this case $\Gamma_2$ is also reducible. 
More precisely, if $(M,N) \in \Gamma_1$ is a redex of a rewrite rule, then
$(N, V) \in \Gamma_2$, for some $V$, is also a redex of the same rewrite rule.
Note that since $\Gamma_1$ and $\Gamma_2$ are consistent, $M$,$N$ and $V$
are all of the same type of syntactic expressions. We show here the case with encrypted
redices, the other case is trivial. 
So suppose that $(\enc R T, \enc U V) \in \Gamma_1$ and $(\enc U V, \enc X Y) \in \Gamma_2.$
Let $\Gamma_1' = \Gamma_1 \setminus \{(\enc R T, \enc U V)\}$
and $\Gamma_2' = \Gamma_2 \setminus \{(\enc U V, \enc X Y)\}.$
Suppose that the following rewrite rule is applied on $\Gamma_1$:
$$
\Gamma_1', (\enc R T, \enc U V) \longrightarrow \Gamma_1', (R, U), (T, V),
$$
and $\Gamma_1 \vdash T \eqm V.$ This entails that $\pi_1(\Gamma_2) \vdash V$ (since
$\pi_2(\Gamma_1) = \pi_1(\Gamma_2)$) and by Definition~\ref{def:theory-consistency}(2),
$\Gamma_2 \vdash V \eqm Y$, so $\Gamma_2$ is indeed reducible. 
The converse, i.e., if $\Gamma_2$ is reducible then $\Gamma_1$ is reducible, can
be proved analogously.

Applying Lemma~\ref{lm:trans-eq} to $\Gamma_1\vdash T \eqm V$
and $\Gamma_2 \vdash V \eqm Y$ obtained above, 
we have $\Gamma_1 \circ \Gamma_2 \vdash T \eqm Y$.
Therefore we can perform the following rewrite:
$$
\Gamma_1 \circ \Gamma_2 = 
\Gamma_1'\circ \Gamma_2', (\enc R T, \enc X Y) 
\longrightarrow
\Gamma_1'\circ \Gamma_2', (R, X), (T, Y)
$$
which contradicts the fact that $\Gamma_1 \circ \Gamma_2$ is irreducible.
Therefore it must be the case that both $\Gamma_1$ and $\Gamma_2$ are irreducible.
\qed
\end{proof}

\begin{lemma}
\label{lm:irreducible-comp}
Let $\Gamma_1$ and $\Gamma_2$ be consistent theories such that $\Gamma_1$ is left-composable
to $\Gamma_2.$ Then $\irred {\Gamma_1}$ is left-composable with $\irred {\Gamma_2}$
and 
$$\irred{(\Gamma_1 \circ \Gamma_2)} = (\irred {\Gamma_1}) \circ 
(\irred{\Gamma_2}).
$$
\end{lemma}
\begin{proof}
We first apply the rewrite rules to $\Gamma_1 \circ \Gamma_2$ until it reaches
its irreducible form. By Lemma~\ref{lm:reduced-comp1}, 
we have $\Gamma_1'$ and $\Gamma_2'$ such that 
$\irred{(\Gamma_1 \circ \Gamma_2)} = \Gamma_1' \circ \Gamma_2'$ and
that $\Gamma_1 \longrightarrow^* \Gamma_1'$ and
$\Gamma_2 \longrightarrow^* \Gamma_2'$. By Lemma~\ref{lm:reduced-comp2}
we have that both $\Gamma_1'$ and $\Gamma_2'$ are irreducible, and since irreducible
forms are unique, it must be the case that 
$\irred{\Gamma_1} = \Gamma_1'$ and $\irred{\Gamma_2} = \Gamma_2'$, and therefore
we have
$$\irred{(\Gamma_1 \circ \Gamma_2)} = (\irred {\Gamma_1}) \circ 
(\irred{\Gamma_2}).
$$
\qed
\end{proof}

\begin{lemma}
\label{lm:unique-projection}
Let $\Gamma$ be a consistent theory. If $\pi_1(\Gamma) \vdash M$
($\pi_2(\Gamma) \vdash M$) then there exists a unique $N$ such
that $\Gamma \vdash M \eqm N$ (respectively, $\Gamma \vdash N \eqm M$).
\end{lemma}
\begin{proof}
By induction on the length of derivations, we can show that
if $\pi_1(\Gamma) \vdash M$ ($\pi_2(\Gamma) \vdash M$)
then there exists an $N$ such that $\Gamma \vdash M \eqm N$ 
(respectively, $\Gamma \vdash N \eqm M$). 
The uniqueness of $N$ follows
immediately from Definition~\ref{def:theory-consistency} (3).
\qed
\end{proof}

\begin{lemma}
\label{lm:unique-decomp}
Let $\Gamma_1$ and $\Gamma_2$ be consistent theories such that $\Gamma_1$
is left-composable to $\Gamma_2.$ If $\Gamma_1 \circ \Gamma_2 \vdash
M \eqm N$, then there exists a unique $R$ such that
$\Gamma_1 \vdash M \eqm R$ and $\Gamma_2 \vdash R \eqm N.$
\end{lemma}
\begin{proof}
Since consistency and composability (of consistent theories) 
are preserved by reduction (Lemma~\ref{lm:reduced-consistent} and 
Lemma~\ref{lm:irreducible-comp}),
without loss of generality, we can assume that $\Gamma_1$ and
$\Gamma_2$ are irreducible, and therefore $\Gamma_1\circ \Gamma_2$
is irreducible as well. So suppose that $\Gamma_1\circ\Gamma_2 \vdash
M \eqm N$. Since $\Gamma_1\circ \Gamma_2$ is irreducible, the derivation
of $\Gamma_1\circ \Gamma_2 \seqsym M \eqm N$ 
does not make use of the left-rules ($el$ and $pl$). 
$R$ can be then constructed inductively by induction on the length
of the derivation and its uniqueness property will follow from
the consistency of $\Gamma_1$ and $\Gamma_2$.
\qed 
\end{proof}

\begin{lemma}
\label{lm:comp-consistency}
Let $\Gamma_1$ and $\Gamma_2$ be consistent theories such that $\Gamma_1$
is left-composable to $\Gamma_2.$ Then $\Gamma_1 \circ \Gamma_2$
is consistent.
\end{lemma}
\begin{proof}
We show that $\Gamma_1\circ \Gamma_2$ satisfies the properties of
consistency defined in Definition~\ref{def:theory-consistency}.
Suppose $\Gamma_1 \circ \Gamma_2 \vdash M \eqm N$. 
By Lemma~\ref{lm:unique-decomp}, there exists a unique $R$
such that $\Gamma_1\vdash M \eqm R$  and $\Gamma_2 \vdash R \eqm N.$
The three properties in Definition~\ref{def:theory-consistency}
are proved as follows:
\begin{enumerate}
\item {\em $M$ and $N$ are of the same type of expressions.}
This trivially holds since $M$, $N$ and $R$ are 
of the same type of expressions by the consistency of $\Gamma_1$
and $\Gamma_2$.

\item {\em If $M = \enc {M_1}{M_2}$ and $N = \enc{N_1}{N_2}$ then
$\pi_1(\Gamma_1 \circ \Gamma_2) \vdash M_2$ implies
$\Gamma_1 \circ \Gamma_2 \vdash M_2 \eqm N_2$, and
$\pi_2(\Gamma_1 \circ \Gamma_2) \vdash N_2$ implies
$\Gamma_1 \circ \Gamma_2 \vdash M_2 \eqm N_2$.
}
We show the first part of the conjunction; the other part is proved
symmetrically. Note that $R = \enc {R_1}{R_2}$,
for some $R_1$ and $R_2$.
Now assume that $\pi_1(\Gamma_1\circ \Gamma_2) \vdash M_2$.
Then $\pi_1(\Gamma_1) \vdash M_2$, hence 
$\Gamma_1 \vdash M_2 \eqm R_2$ by the consistency of $\Gamma_1$.
From this, it follows that $\pi_1(\Gamma_2) \vdash R_2$ and
therefore $\Gamma_2 \vdash R_2 \eqm N_2$ by the consistency of 
$\Gamma_2$. By Lemma~\ref{lm:trans-eq}, this means that
$\Gamma_1 \circ \Gamma_2 \vdash M_2 \eqm N_2$ as required.

\item {\em For any $T$, $\Gamma_1 \circ \Gamma_2 \vdash M \eqm T$
implies $T = N$ and $\Gamma_1 \circ \Gamma_2 \vdash T \eqm N$
implies $T = M$.}
We show the first case; the other is symmetric.
Suppose $\Gamma_1 \circ \Gamma_2\vdash M \eqm T$.
By Lemma~\ref{lm:unique-decomp}, there exists a unique $U$ such
that $\Gamma_1 \vdash M \eqm U$ and $\Gamma_2 \vdash U \eqm T$.
But this means $U = R$, by the consistency of $\Gamma_1$,
and $T = N$, by the consistency of $\Gamma_2.$
\end{enumerate}
\qed
\end{proof}

\section{Open bisimulation}
\label{sec:open}

Open bisimulation for the spi-calculus to be presented in this section 
is similar to other environment-sensitive bisimulations, in the sense that 
it is also indexed by some structure representing the knowledge of the environment. 
A candidate for representing this knowledge is the observer theory
presented earlier. However, since the crucial feature of open bisimulation
is the symbolic representation of input values, extra structures need to 
be added to observer theories to capture dependencies between various
symbolic input values at different stages of bisimulation checking. 
The notion of {\em symbolic traces} as defined in \cite{boreale01icalp} 
conveniently captures this sort of dependency. Open bisimulation
is indexed by pairs of a variant of symbolic traces,
called {\em bi-traces}. The important properties we need to establish
regarding bi-traces are that they can be soundly interpreted as
observer theories, and they behave well with respect to substitutions of
input values. 

In the following, we use the notation $[x_1,\ldots,x_n]$ to denote
a list whose elements are $x_1, \ldots,x_n.$ The empty list
is denoted by $[\,].$
Concatenation of a list $l_1$ with another list $l_2$ 
is denoted with $l_1.l_2$, if $l_2$ is appended to
the end of $l_1$.
If $l_2$ is a singleton list, say $[x]$, 
then we write $l_1.x$ instead of $l_1.[x],$
likewise $x.l_1$ instead of $[x].l_1.$



\begin{definition}
\label{def:bi-traces}
An {\em I/O pair} is a pair of messages marked with
$i$ (indicating input) or $o$ (indicating output),
i.e., it is of the form $(M,N)^i$ or
$(M,N)^o.$
A {\em bi-trace} is a {\em list} of I/O message pairs, ranged
over by $h$. We denote with $\pi_1(h)$ the list obtained
from $h$ by taking the first component of the pairs in $h$.
The list $\pi_2(h)$ is defined analogously.
Bi-traces are subject to the following restriction: 
if $h = h_1.(M,N)^o.h_2$ then $\freeFN{M,N} \subseteq \freeFN{h_1}.$
If $h$ is 
$$
[(M_1,N_1)^{l_1}, \ldots, (M_k, N_k)^{l_k}]
$$
then the {\em inverse} of $h$, written $h^{-1}$,
is the list 
$$
[(N_1,M_1)^{l_1},\ldots,(N_k,M_k)^{l_k}].
$$
We write $\{h\}$ to denote the set
$$
\{(M,N) \mid (M,N)^i \in h \hbox{ or } (M,N)^o \in h\}.
$$
\end{definition}

The underlying idea in the bi-trace representation is that
{\em names are symbolic values}. This explains the requirement
that the free names of an output pair in a bi-trace 
must appear before the output pair. In other words, input values
(i.e., names) are created only at input pairs. 

Given a bi-trace $h$, the underlying set $\{h\}$ is obviously 
an observer theory. Application of a substitution pair $(\theta_1,\theta_2)$
to a bi-trace is defined element-wise, i.e.,
$$
\begin{array}{ll}
[\,](\theta_1,\theta_2) & =  [\,]\\
((M,N)^*.h') (\theta_1,\theta_2) & = (M\theta_1,N\theta_2)^*.(h'(\theta_1,\theta_2))\\
\end{array}
$$
where $*$ is either $i$ or $o.$
Bi-traces are essentially theories with added structures. As such, we also associate
a notion of consistency with bi-traces. 
As in Boreale's symbolic traces ~ \cite{boreale01icalp}, bi-traces consistency needs to
take into account the fact that their instantiations correspond to concrete traces.
Not all instantiations of symbolic traces give rise to correct concrete traces. 
For example, the processes
$
P = a(x).(\nu k) \bar a k.\bar a x.
$
has a symbolic trace $a x.\bar a k.\bar a x,$ but instantiating
$x$ to $k$ produces a concrete trace
$a k.\bar a k.\bar a k,$ which does not correspond to any actual
trace the process $P$ can produce, since the input $x$ happens before
$k$ is extruded. Consistency conditions for bi-traces are more complicated
than symbolic traces, since we need extra conditions ensuring the consistency
of the observer theory underlying the traces. 
We first define a notion of respectful substitutions for bi-traces.
In the following we shall write $h \vdash M \eqm N$, instead of a more type-correct
version $\{h\} \vdash M \eqm N$, when we consider an equivalent 
pair of messages under the theory obtained from a bi-trace $h$.

\begin{definition}
\label{def:respectful-subst}
A substitution pair $\vec \theta = (\theta_1,\theta_2)$ respects a 
bi-trace $h$ if whenever $h = h_1.(M,N)^i.h_2$,  then 
for every $x \in \freeFN{M,N}$ it holds that
$$
h_1\vec\theta \vdash x\theta_1 \eqm x\theta_2.
$$
\end{definition}

The requirement that every input pair be deducible from its predecessors
in the bi-trace captures the dependency of the names of the input pair
on their preceding input/output pairs, and thus avoids unsound instantiations
as described above. 
At this point, it is instructive to examine the case where the elements
of bi-traces are pairs of names or rigid names. Consider for example
the bi-trace
$$
(x,x)^i.(\abf, \abf)^o.(y,y)^i.(\bbf,\bbf)^o.
$$
There is a respectful substitution that identifies $x$ and $y$, or
$y$ with $\abf$, but there are no respectful substitutions that identify
$x$ with $\abf$, $y$ with $\bbf$ nor $\abf$ with $\bbf$.
Thus this bi-trace captures a restricted notion of distinction~\cite{sangiorgi96acta}.
Rigid names encodes an implicit distinction: no two rigid names can
be identified by substitutions, whereas the position of names encode their
respective scopes.

We now proceed to defining bi-trace consistency.
\begin{definition}
\label{def:bi-trace-consistent}
We define the notion of {\em consistent bi-traces} inductively on the length
of bi-traces as follows:
\begin{enumerate}
\item The empty bi-trace is consistent.
\item If $h$ is a consistent bi-trace then $h.(M,N)^i$ is also a consistent
bi-trace, provided that $h \vdash M \eqm N$.
\item If $h$ is a consistent bi-trace, then $h' = h.(M,N)^o$ is a consistent bi-trace,
provided that for every $h$-respectful substitution pair 
$\vec \theta$, if $h\vec \theta$ is a consistent bi-trace
then $\{h'\vec \theta \}$ is a consistent theory.
\end{enumerate}
\end{definition}

Note that in item (3) in the above definition, there is a negative occurence
of consistent bi-traces. But since this occurence is about a smaller trace,
it is already defined by induction, and therefore the definition is still 
well-founded. 
In the same item we quantify over all respectful substitutions. This
is unfortunate from the viewpoint of bisimulation checking but it is
unavoidable if we want the notion of consistency to be closed under
respectful substitutions. Consider the following example: let
$h$ be the bi-trace:
$$
(\abf,\abf)^o.(\bbf,\bbf)^o.(x,x)^i.(\enc x \kbf, \enc \abf \kbf)^o.(\enc \bbf \kbf, \enc x \kbf)^o.
$$
If we drop the quantification on respectful substitutions, then 
this trace would be considered consistent. However, under the
respectful substitution pair $([\bbf/x] , [\bbf/x]),$
the above bi-trace will be instantiated to
$$
(\abf,\abf)^o.(\bbf,\bbf)^o.(\bbf,\bbf)^i.(\enc \bbf \kbf, \enc \abf \kbf)^o.(\enc \bbf \kbf, \enc \bbf \kbf)^o
$$
which gives rise to an inconsistent theory.
Complete finite characterisation of consistent bi-traces is left for
future work.

Note that for any given a bi-trace $h$, the empty substitution pair 
$(\epsilon,\epsilon)$ is obviously an $h$-respectful substitution. 

\subsection{Properties of bi-traces}

We now look at some properties of bi-traces. Among the important ones
are those that concern {\em composition} of bi-traces. 

\begin{definition}
\label{def:comp-bitraces}
{\em Composition of bi-traces.}
Two bi-traces can be composed if they have the same length and match
element wise. More precisely, given two bi-traces
$$
h_1 = [(R_1,T_1)^{p_1}, \cdots, (R_m, T_m)^{p_m}]
$$
$$
h_2 = [(U_1,V_1)^{q_1}, \cdots, (U_n, V_n)^{q_n}]
$$
we say $h_1$ is {\em left-composable} to $h_2$ (equivalently, $h_2$ is
{\em right-composable} to $h_1$) if and only if $m = n$ and $T_k = U_k$ and $p_k = q_k$ for every
$k \in \{1,\dots, n\}$. Their composition, written $h_1 \circ h_2$, is
$$
h_1 \circ h_2 = [(R_1, V_1)^{p_1}, \cdots, (R_m, V_m)^{p_m}]
$$
\end{definition}

Note that there is a subtle difference between composability of bi-traces
and theories. In Definition~\ref{def:comp-bitraces} we do not require that
$T_1,\dots,T_m$ (likewise, $U_1,\dots,U_n$) are pairwise distinct messages,
since their positions in the list determine uniquely the composition.
So in general, compositions of bi-traces need not coincide with 
compositions of their underlying theories. They do coincide, however, if
we restrict to consistent bi-traces.

\begin{lemma}
\label{lm:subtrace-consistent}
If $h = h_1.h_2$ is a consistent bi-trace then so is $h_1$.
\end{lemma}

\begin{lemma}
\label{lm:respectful-subst}
Let $h$ be a bi-trace. If $\vec \theta = (\theta_1,\theta_2)$
respects $h$, then for every name $x \in \freeFN{h}$,
we have $h\vec \theta \vdash x\theta_1 \eqm x\theta_2.$
\end{lemma}
\begin{proof}
The proof is by induction on the length of $h$. 
The case with $h=[]$ is trivial. We look at the other two cases:
\begin{itemize}
\item Suppose $h = h'.(M,N)^i$. Since $\vec \theta$ also respects $h'$,
by the induction hypothesis we have for every $y \in \freeFN{h'}$,
$h'\vec \theta \vdash y\theta_1 \eqm y\theta_2$, and by the monotonicity
of $\vdash$, we have $h\vec \theta \vdash y\theta_1 \eqm y\theta_2.$
For every name $z \in \freeFN{M,N} \setminus \freeFN{h'}$, we also
have $h\vec \theta \vdash z\theta_1 \eqm z\theta_2,$ since
$\vec \theta$ respects $h$. Therefore for every name $x\in \freeFN{h}$
we indeed have $h\vec \theta \vdash x\theta_1 \eqm x\theta_2.$
\item Suppose $h=h'.(M,N)^o$. By the restriction on bi-traces,
it must be the case that $\freeFN{M,N} \subseteq \freeFN{h'}$,
therefore $\freeFN{h} = \freeFN{h'}$. Therefore by induction hypothesis
we have that for every $x \in \freeFN{h}$, $h\vec\theta \vdash x\theta_1 \eqm
x\theta_2.$
\end{itemize}
\qed
\end{proof}

\begin{lemma}
\label{lm:respectful-subst2}
Let $h =  h'.(M,N)^i$ be a bi-trace and let $\vec \theta = 
(\theta_1,\theta_2)$ be an $h$-respectful substitution. Then
$h'\vec\theta \vdash x\theta_1 \eqm x\theta_2,$ for every $x \in \freeFN{h}.$
\end{lemma}
\begin{proof}
Applying Lemma~\ref{lm:respectful-subst} to $h'$, 
we have for every $x\in \freeFN{h'}$, $h'\vec \theta \vdash x\theta_1 \eqm x\theta_2.$
Now by Definition~\ref{def:respectful-subst}, we have 
$h'\vec \theta \vdash x\theta_1 \eqm x\theta_2$ for every $x \in \freeFN{M,N}.$
We therefore have covered all the free names in $h$.
\qed
\end{proof}

\begin{lemma}
\label{lm:respectful-subst-comp}
Let $h$ be a consistent bi-trace, let $\vec\theta = (\theta_1,\theta_2)$ be an
$h$-respectful substitution pair, and let $\vec\gamma = (\gamma_1,\gamma_2)$
be an $h\vec\theta$-respectful substitution pair. Then $\vec\theta \circ \vec\gamma$
is also an $h$-respectful substitution pair.
\end{lemma}
\begin{proof}
We have to show that whenever $h = h_1.(M,N)^i.h_2$,
for every $x \in \freeFN{M,N}$, $(h_1\vec \theta)\vec \gamma \vdash
(x\theta_1)\gamma_1 \eqm (x\theta_2)\gamma_2.$
Since $\vec \theta$ respects $h$ and $\vec \gamma$ respects $h\vec \theta$, 
we have that
$$
\hbox{for every $x' \in \freeFN{M,N}$, } h_1\vec \theta \vdash x'\theta_1 \eqm x'\theta_2,
$$
$$
\hbox{for every $y \in \freeFN{M\theta_1, N\theta_2}$, }
(h_1\vec\theta)\vec\gamma \vdash y\gamma_1 \eqm y\gamma_2.
$$
Now since $x\in \freeFN{M,N}$, it follows that 
$\freeFN{x\theta_1, x\theta_2} \subseteq \freeFN{M\theta_1,N\theta_2}.$
From Lemma~\ref{lm:respectful-subst2}, we have 
$$
h_1\vec\theta\vec\gamma \vdash y\gamma_1 \eqm y\gamma_2
$$
for every $y \in \freeFN{h_1\vec\theta, M\theta_1,N\theta_2}.$
Therefore, we can apply Lemma~\ref{lm:clo-subst2} to get
$(h\vec\theta)\vec \gamma \vdash (x\theta_1)\gamma_1 \eqm (x\theta_2)\gamma_2.$
\qed
\end{proof}

\begin{lemma}
\label{lm:respect-preserves-consistency}
If $h$ is a consistent bi-trace and $\vec \theta = (\theta_1,\theta_2)$
respects $h$, then $h \vec \theta$ is also a consistent bi-trace.
\end{lemma}
\begin{proof}
The proof is by induction on the length of $h$.
The base case is obvious. There are two inductive cases:
Suppose $h = h'.(M,N)^i$. Since $\vec \theta$ respects $h'$,
by the induction hypothesis we know that $h'\vec\theta$ is consistent.
We have to show that $h'\vec\theta \vdash M\theta_1 \eqm N\theta_2$.
From Lemma~\ref{lm:respectful-subst} and Definition~\ref{def:respectful-subst},
it follows that for every $x \in \freeFN{h}$, 
$h'\vec \theta \vdash x\theta_1 \eqm x\theta_2.$
Therefore by Lemma~\ref{lm:clo-subst2},
we have $h'\vec\theta \vdash M\theta_1 \eqm N\theta_2$ as required.

Suppose $h=h'.(M,N)^o.$ Since $h$ is consistent, we have that for
every $h'$-respectful substitution pair $\vec \sigma = (\sigma_1,\sigma_2)$
(including $\vec \theta$), if $h'\vec \sigma$ is a consistent bi-trace
then $\{h\vec\sigma\}$ is a consistent theory. By the induction hypothesis,
$h'\vec \sigma$ is consistent, and therefore
$\{h\vec \sigma\}$ is a consistent theory, for every respectful $\vec \sigma$.
The statement we want to prove is the following:
for every $h'\vec \theta$-respectful substitution pair 
$\vec \gamma = (\gamma_1,\gamma_2),$ if $(h'\vec \theta)\vec \gamma$
is a consistent bi-trace, then $\{(h'\vec \theta)\vec \gamma \}.$
It is enough to show that $\vec \theta \circ \vec \gamma$ is an
$h'$-respectful substitution pair, which follows from Lemma~\ref{lm:respectful-subst-comp}.
\qed
\end{proof}

\begin{lemma}
\label{lm:bi-trace-to-theory}
If $h$ is a consistent bi-trace then $\{h\}$ is a consistent theory.
\end{lemma}

\begin{lemma}
If $h$ is consistent then so is $h^{-1}$.
\end{lemma}

\begin{lemma}
Let $h_1$ and $h_2$ be two consistent bi-traces such that $h_1$ is left-composable
with $h_2$. Then $\{h_1\}$ is left composable to $\{h_2\}$ and
$\{h_1\} \circ \{h_2\} = \{h_1 \circ h_2 \}.$
\end{lemma}

\begin{lemma}
\label{lm:names-in-bi-traces}
Let $h$ be a consistent bi-trace. Then $\freeFN{\pi_1(h)} = \freeFN{\pi_2(h)}.$
\end{lemma}

The following lemma is crucial to the proof of transitivity of open bisimulation.

\begin{lemma}
\label{lm:separating-subst}
Let $h_1$ and $h_2$ be consistent and composable bi-traces such
that $h_1 \circ h_2$ is also consistent.
Let $(\theta_1,\theta_2)$ be a substitution pair that respects 
$h_1\circ h_2.$ 
Then there exists a substitution $\rho$ such that
$(\theta_1,\rho)$ respects $h_1$ and $(\rho,\theta_2)$ respects $h_2$.
\end{lemma}
\begin{proof}
We construct $\rho$ by induction on the length of $h_1\circ h_2.$
At each stage of the induction, we construct a substitution $\rho$
satisfying the statement of the lemma.
In the base case, where $h_1\circ h_2$ is the empty list, we
take $\rho$ to be the empty substitution. The inductive cases
are handled as follows.
\begin{itemize}
\item $h_1 = h_1'.(M,N)^i$ and $h_2 = h_2'.(N,R)^i$.
By the induction hypothesis, there is a substitution $\rho'$ such that
$(\theta_1,\rho')$ respects $h_1'$ and $(\rho',\theta_2)$ respects
$h_2'.$ 
We will make use of the following facts:
\begin{itemize}
\item $h_1'$ and $h_2'$ are consistent, and since $(\theta_1,\rho')$
respects $h_1'$ and $(\rho', \theta_2)$ respects $h_2'$, it follows
from Lemma~\ref{lm:respect-preserves-consistency} that 
$h_1'(\theta_1,\rho')$ and $h_2'(\rho',\theta_2)$ are also consistent.
\item $(h_1'\circ h_2')\vec \theta = (h_1'(\theta_1,\rho')) \circ 
(h_2'(\rho',\theta_2)).$ 
\item $(h_1'\circ h_2')$ is consistent and therefore, by 
Lemma~\ref{lm:respect-preserves-consistency},
$(h_1'\circ h_2')\vec \theta$ is consistent a bi-trace and
its underlying theory is also consistent (Lemma~\ref{lm:bi-trace-to-theory}).
\item Since $\vec \theta$ respects $h_1\circ h_2$, 
by Lemma~\ref{lm:respectful-subst2},
we have that for every $x \in \freeFN{h_1 \circ h_2}$,
$(h_1'\circ h_2')\vec \theta \vdash x\theta_1 \eqm x\theta_2.$
\end{itemize}
From these facts, and Lemma~\ref{lm:unique-decomp}, 
for every $x \in \freeFN{h_1,h_2}$, there exists a unique
$U$ such that $h_1'(\theta_1,\rho') \vdash x\theta_1 \eqm U$
and $h_2'(\rho',\theta_2) \vdash U \eqm x\theta_2.$
We let $f(x)$ denote the unique $U$ obtained this way. 
Now define $\rho$ as follows:
$$
\rho(x) = 
\left\{
\begin{array}{ll}
\rho'(x), & \hbox{if $x\in\freeFN{h_1',h_2'}$,} \\
f(x), & \hbox{if $x \in \freeFN{h_1,h_2}$ but $x\not \in \freeFN{h_1',h_2'}$,}\\
x, & \hbox{otherwise.}
\end{array}
\right.
$$
Note that by Lemma~\ref{lm:names-in-bi-traces}, 
$\freeFN{h_1',h_2'} = \freeFN{h_1'} = \freeFN{h_2'}.$
We now show that $(\theta_1,\rho)$ respects $h_1$ and 
$(\rho,\theta_2)$ respects $h_2$.
\begin{enumerate}
\item $(\theta_1,\rho)$ respects $h_1$: Since $\rho$ and $\rho'$ coincide
on $\freeFN{h_1'}$, $(\theta_1, \rho)$ also respects $h_1'$. We therefore need only to 
check that $h_1'(\theta_1,\rho) \vdash x\theta_1 \eqm x\rho$,
for every $x \in \freeFN{M,N}\setminus \freeFN{h_1'}.$
This follows immediately from the construction of 
$x\rho$ discussed above.
\item $(\rho, \theta_2)$ respects $h_2$: symmetric to the previous case. 
\end{enumerate}

\item $h_1 = h_1'.(M,N)^o$ and $h_2 = h_2'.(N,R)^o$. 
In this case, $\freeFN{M,N,R} \subseteq \freeFN{h_1',h_2'}$.
By the induction hypothesis, we have a substitution $\rho'$ such that
$(\theta_1,\rho')$ respects $h_1'$ and $(\rho',\theta_2)$ respects
$h_2'.$ We simply define $\rho = \rho'.$ 
It follows immediately from Definition~\ref{def:respectful-subst}
that $(\theta_1,\rho)$ respects $h_1$ and $(\rho,\theta_2)$ respects
$h_2.$
\end{itemize}
\qed
\end{proof}

\begin{lemma}
\label{lm:trace-comp-consistent}
Let $h_1$ and $h_2$ be consistent bi-traces. Then their composition,
$h_1 \circ h_2$, if defined, is also a consistent bi-trace.
\end{lemma}
\begin{proof}
By induction on the length of $h_1\circ h_2.$ The base case is obvious.
The inductive cases are handled as follows:
\begin{itemize}
\item $h_1 = h_1'.(M,N)^i$ and $h_2 = h_2'.(N,R)^i$:
By induction hypothesis $h_1'\circ h_2'$ is consistent. 
Since $h_1$ and $h_2$ are consistent, we have that
$h_1'\vdash M \eqm N$ and $h_2'\vdash N \eqm R$, and applying
Lemma~\ref{lm:trans-eq}, we have $h_1'\circ h_2' \vdash M \eqm N.$ 
Therefore $h_1\circ h_2$ is consistent.
\item $h_1 = h_1'.(M,N)^o$ and $h_2 = h_2'.(N,R)^o$:
By induction hypothesis $h_1'\circ h_2'$ is consistent. We need to show
that for every $(h_1'\circ h_2')$-respectful substitution pair 
$\vec\theta = (\theta_1,\theta_2)$,
if $(h_1'\circ h_2')\vec \theta$ is a consistent bi-trace then
$\{(h_1\circ h_2)\vec \theta\}$ is a consistent theory.
So let us suppose that $(h_1'\circ h_2')\vec \theta$ is consistent.
From Lemma~\ref{lm:separating-subst}, there exists a substitution $\rho$ such that
$(\theta_1,\rho)$ respects $h_1'$ and $(\rho,\theta_2)$ respects $h_2'$.
And since $\freeFN{M,N} \subseteq \freeFN{h_1'}$ and
$\freeFN{N,R} \subseteq \freeFN{h_2'}$, we have
$(\theta_1,\rho)$ respects $h_1$ and $(\rho,\theta_2)$ respects $h_2$.
Therefore, by Lemma~\ref{lm:respect-preserves-consistency}, 
$h_1(\theta_1,\rho)$ and $h_2(\rho,\theta_2)$ are consistent bi-traces.
Since $(h_1\circ h_2)\vec \theta = (h_1(\theta_1,\rho)) \circ (h_2(\rho,\theta_2))$,
and therefore 
$\{ (h_1\circ h_2)\vec \theta \} = 
\{(h_1(\theta_1,\rho))\} \circ\{ (h_2(\rho,\theta_2))\}$
it follows from Lemma~\ref{lm:comp-consistency} that $\{ (h_1\circ h_2) \vec \theta \}$
is indeed a consistent theory.
\end{itemize}
\qed
\end{proof}

\subsection{Definition of open bisimulation}

\begin{definition}
A {\em traced process pair} is a triple $(h,P,Q)$ where
$h$ is a bi-trace, $P$ and $Q$ are processes such that
$\freeFN{P,Q} \subseteq \freeFN{h}.$ 
Let ${\cal R}$ be a set of traced process pairs.
We write $h \vdash P~{\cal R}~Q$ to denote
the fact that $(h,P,Q) \in {\cal R}.$
${\cal R}$ is consistent if for every $h \vdash P ~ {\cal R} ~ Q$,
$h$ is consistent.
The inverse of ${\cal R}$, written ${\cal R}^{-1}$, is
the set
$$
\{(h^{-1},Q,P) \mid (h,P,Q) \in {\cal R} \}.
$$
${\cal R}$ is {\em symmetric} if ${\cal R} = {\cal R}^{-1}.$
\end{definition}

\begin{definition}
A bi-trace $h$ is called a {\em universal bi-trace} if $h$ consists only of
input-pairs of names, i.e., it is of the form
$(x_1,x_1)^i.\cdots.(x_n,x_n)^i$, where each $x_i$ is a name. 
\end{definition}

\begin{definition}
\label{def:open-bisim}
{\em Open bisimulation.}
A set of traced process pairs ${\cal R}$ is a {\em strong open bisimulation}
if ${\cal R}$ is consistent and symmetric, and 
if $h \vdash P~{\cal R}~Q$ then 
for all substitution pair $\vec \theta = (\theta_1,\theta_2)$
that respects $h$, the following hold:
\begin{enumerate}
\item If $\one{P\theta_1}{\tau}{P'}$ then there exists $Q'$
such that $\one{Q\theta_2}{\tau}{Q'}$ and
$h\vec \theta \vdash P'~{\cal R}~Q'.$
\item If $\one{P\theta_1}{M}{(x)P'}$, where 
$x \not \in \freeFN{h\vec \theta},$
and $\pi_1(h\vec\theta) \vdash M$ 
then there exists $Q'$ such that 
$\one{Q\theta_2}{N}{(x)Q'}$ 
and 
$$
h\vec\theta . (M,N)^i.(x,x)^i \vdash P'~{\cal R}~Q'.
$$
\item If $\one{P\theta_1}{\bar M}{(\nu \vec x)\langle M' \rangle P'},$
and $\pi_1(h\vec\theta) \vdash M$
then there exist $N$, $N'$ and $Q'$ such that $\one{Q\theta_2}{\bar N}{(\nu \vec y)\langle N' \rangle Q'},$
and
$$
h\vec\theta . (M, N)^{i}.(M'[\vec \cbf/\vec x],N'[\vec {\dbf}/\vec y])^o 
\vdash P'[\vec {\cbf}/\vec x] ~{\cal R}~ Q'[\vec {\dbf}/\vec y],
$$
where $\{ \vec c, \vec d \} \cap \freeRN{h\vec\theta,P\theta_1,Q\theta_2} = \emptyset.$ 
\end{enumerate}
We denote with $\obisim$ the union of all open bisimulations. 
We say that $P$ and $Q$ are {\em strong open $h$-bisimilar}, 
written $P \sim^h_o Q$, if $(h,P,Q) \in \ \obisim.$ 
They are said to be {\em strong open bisimilar}, written $P \sim_o Q$, if
$\freeRN{P,Q} = \emptyset$ and $P \sim^h_o Q$ for a universal bi-trace $h$.
\end{definition}

Notice that strong open bisimilarity $\sim_o$ is defined on pure processes, i.e.,
those processes without free occurrences of rigid names.

\begin{lemma}
The relation $\obisim$ is a strong open bisimulation.
\end{lemma}

\section{Up-to techniques}
\label{sec:upto}

We define several up-to techniques for open bisimulation.
The main purpose of these techniques is to prove congruence results for
open bisimilarity, in particular, closure under parallel composition,
and to prove soundness of open bisimilarity with respect to testing equivalence.
Up-to techniques are also useful in checking bisimulation since in
certain cases it allows one to finitely demonstrate bisimilarity of processes.
The proof techniques used in this section derive mainly from the work
of Boreale et. al.~\cite{boreale02sjc}. We first need to introduce
several notions, parallel to those in \cite{boreale02sjc}, and adapting
their up-to techniques to open bisimulation.

It is quite well-known that open bisimilarity is not closed under parallel composition
with arbitrary processes, since these extra processes might introduce inconsistency
into the observer theory or may reveal other knowledge that causes the composed
processes to behave differently. For example, it can be shown that
$$
(\enc \abf \kbf, \enc \abf \kbf)^o.(x,x)^i \vdash [x = \abf] \bar \abf x.0 \obisim 0,
$$
since $\abf$ is encrpyted with the key $\kbf$ which is unknown to the 
observer, which means that the observer cannot possibly feed $\abf$ into the input
$x.$ Thus the match prefix in the process $[x = \abf]\bar \abf x.0$ will 
evaluate to true and the process is stuck. However, if we put the processes in
paralle with $\bar x \kbf$, the composed processes become
$$
[x = \abf] \bar \abf x.0 ~|~ \bar x \kbf
\qquad
\hbox{ and }
\qquad
0 ~|~ \bar x \kbf.
$$
Both processes can output $\kbf$ on $x$, leading to the bi-trace
$$
(\enc \abf \kbf, \enc \abf \kbf)^o.(x,x)^i.(\kbf, \kbf)^o
$$
at which point, the observer can decrypt the first output pair to get to
$\abf$, and under this knowledge, $[x = \abf] \bar \abf x.0$ is no longer
bisimilar to $0$.

Given the above observeration, in defining closure under parallel composition,
we need to make sure that the processes we are composing with do not reveal or add
any extra information for the observer. A way to do this is to restrict the 
composition to processes obtained by instantiating pure processes with the current
knowledge of the observer. This is defined via a notion of equivalent substitutions,
given in the following.

\begin{definition}
\label{def:equiv-subst}
Let $h$ be a consistent bi-trace. Given two substitutions  
$\theta_1$ and $\theta_2$, we say that $\theta_1$ is $h$-equivalent
to $\theta_2$, written $\theta_1 \eqm_h \theta_2$, 
if $\dom {\theta_1} = \dom {\theta_2}$ and 
for every $x \in \dom{\theta_1}$, we have 
$h \vdash x\theta_1 \eqm x\theta_2$
and $\freeFN{x\theta_1,x\theta_2} \subseteq \freeFN{h}.$
A substitution $\sigma$ {\em extends} $\theta$, written
$\theta \preceq \sigma$, if
$\sigma(x) = \theta(x)$ for every $x \in \dom{\theta}.$
\end{definition}

\begin{lemma}
\label{lm:comp-subst}
Let $h$ be a consistent bi-trace, let $\vec \theta = (\theta_1,\theta_2)$
be an $h$-respectful substitution and let $\sigma_1$ and $\sigma_2$
be substitutions such that $\sigma_1 \eqm_h \sigma_2.$
Let $\sigma_1'$ and $\sigma_2'$ be the following substitutions:
$$
\sigma_1' = \rsubst{(\sigma_1 \circ \theta_1)}{\dom{\sigma_1}}
\qquad \hbox{ and } \qquad
\sigma_2' = \rsubst{(\sigma_2 \circ \theta_2)}{\dom{\sigma_2}}.
$$
Then $\sigma_1' \eqm_{h\vec\theta} \sigma_2'.$
\end{lemma}
\begin{proof}
We have to show that $h\vec\theta \vdash x\sigma_1\theta_1 \eqm x\sigma_2\theta_2,$
for every $x\in \dom{\sigma_1'}.$
Since we have $h \vdash x\sigma_1 \eqm x\sigma_2$, and since $\vec \theta$ respects $h$
and $\freeFN{x\sigma_1,x\sigma_2} \subseteq \freeFN{h}$, by Lemma~\ref{lm:respectful-subst}
and Lemma~\ref{lm:clo-subst2}, we have $h\vec\theta\vdash x\sigma_1\theta_1 \eqm x\sigma_2\theta_2.$
It remains to show that $\freeFN{x\sigma_1\theta_1,x\sigma_2\theta_2} \subseteq \freeFN{h\vec\theta}.$
But this follows immediately from the fact that $\freeFN{x\sigma_1,x\sigma_2} \subseteq \freeFN{h}.$
\qed
\end{proof}

\begin{lemma}
\label{lm:equiv-subst}
Let $h$ be a consistent bi-trace and let $\sigma_1$ and
$\sigma_2$ be substitutions such that $\sigma_1 \eqm_h \sigma_2.$
Let $M$ and $N$ be messages such that $\freeFN{M,N}\subseteq \dom{\sigma_1}$
and $\freeRN{M,N} = \emptyset.$ Then the following hold:
\begin{enumerate}
\item $h \vdash M\sigma_1 \eqm M \sigma_2.$
\item $M\sigma_1 = N\sigma_1$ if and only if $M\sigma_2 = N\sigma_2.$
\end{enumerate}
\end{lemma}
\begin{proof}
Statement (1) is proved by induction on the size of $M$. Statement (2)
then follows from (1) and the consistency of $h$.
\qed
\end{proof}

Note that item (2) in the above lemma is a simplification of 
the equivalence conditions for substitutions in the work of
Boreale et. al.~\cite{boreale02sjc}. In their work, processes can have
boolean guards, constructed from the standard connectives of classical logic and equality, 
and they show that satisfiability of any formula is preserved under equivalent
substitutions.

The next lemma is crucial to the soundness of up-to parallel composition. It shows that
one-step transitions for pure processes are invariant under equivalent substitutions. 

\begin{lemma}
\label{lm:equiv-subst-trans}
Let $h$ be a consistent bi-trace, let $\sigma_1$ and $\sigma_2$ be
substitutions such that $\sigma_1 \eqm_h \sigma_2$, and let
$R$ be a process such that $\freeFN{R} \subseteq \dom{\sigma_1}$
and $\freeRN{R} = \emptyset$. 
If $\one{R\sigma_1}{M}{R'}$ then there exist $\sigma_1 \preceq \sigma_1'$,
$\sigma_2 \preceq \sigma_2'$, $U$ and $Q$ such
that $\sigma_1' \eqm_h \sigma_2'$, 
$\freeFN{U,Q} \subseteq \dom{\sigma_1'}$,
$\freeRN{U,Q} = \emptyset$, 
$M = U\sigma_1'$, $R' = Q\sigma_1'$ and 
$\one{R\sigma_2}{U\sigma_2'}{Q\sigma_2'}.$
\end{lemma}
\begin{proof}
The proof is by induction on the height of the derivation of
the transition relation $\one{R\sigma_1}{M}{R'}.$
Most cases follow straightforwardly from the induction hypothesis.
The non-trivial cases are those that involve reductions of paired
and encrypted messages. We examine the case with encryptions, 
the other case is treated similarly. 

Suppose $R = \spcase L x N P$ and the transition is derived
as follows:
$$
\infer[]
{\one{\spcase {L\sigma_1} x {N\sigma_1} {P\sigma_1}}{M}{R'} }
{\spcase {L\sigma_1} x {N\sigma_1} {P\sigma_1} > P\sigma_1[L_1/x] & \one{P\sigma_1[L_1/x]}{M}{R'} }
$$
Here we assume, without loss of generality, that $x$ is chosen to be
fresh with respect to $\sigma_1$, $\sigma_2$, $R$ and $h.$
It must be the case that $L\sigma_1 = \enc {L_1}{N\sigma_1}.$ 
Now by Lemma~\ref{lm:equiv-subst} we know that $h\vdash N\sigma_1 \eqm N\sigma_2$
and $h\vdash L\sigma_1 \eqm L\sigma_2.$ Therefore, by Lemma~\ref{lm:right-invertible},
$L\sigma_2$ must also be of the form $\enc {L_2}{N\sigma_1}$ for some 
$L_2$ such that $h\vdash L_1 \eqm L_2.$
Let us extend $\sigma_1$ and $\sigma_2$ to the following substitutions:
$$
\theta_1 = \sigma_1 \cup \{x \mapsto L_1 \}
\hbox{ and }
\theta_2 = \sigma_2 \cup \{x \mapsto L_2 \}.
$$
Obviously, $\theta_1 \eqm_h \theta_2$. Therefore by induction hypothesis,
there exist
$\theta_1 \preceq \theta_1'$, $\theta_2 \preceq \theta_2'$,
$U'$ and $Q'$ such that $\theta_1' \eqm_h \theta_2'$, 
$U'\theta_1 = M$,
$Q'\theta_1' = R'$ and
$\one{P\theta_2}{U'\theta_2}{Q'\theta_2'}.$
We now define $U$ and $Q$ to be $U'$ and $Q'$, respectively,
and let $\sigma_1' = \theta_1'$ and $\sigma_2' = \theta_2'$. 
Obviously, $\sigma_1 \preceq \sigma_1'$, $\sigma_2 \preceq \sigma_2'$
and $\sigma_1' \eqm_h \sigma_2'.$
The transition from $R\sigma_2$ is therefore inferred as follows:
$$
\infer[]
{\one{\spcase {L\sigma_2}{x}{N\sigma_2}{P\sigma_2}}{U\sigma_2'}{Q\sigma_2'}}
{
\spcase {L\sigma_2} {x} {N\sigma_2}{P\sigma_2} > P\theta_2
&
\one{P\theta_2}{U\sigma_2'}{Q\sigma_2'}
}
$$
\qed
\end{proof}

We need a few relations on bi-traces to describe the following up-to rules.

\begin{definition}
The relations $<_i$, $<_o$ and $<_f$ on bi-traces are defined
as follows:
$$
\begin{array}{ll}
\hbox{(weakening)} & h <_w h', \hbox{ if $h = h_1.h_2$ and $h' = h_1.(M,N)^*.h_2$, 
    where $* \in \{i,o\}$ and $\freeFN{M,N} \subseteq \freeFN{h_1}$.}\\
\hbox{(contraction)} & h <_c h', \hbox{ if  $h = h_1.(M,N)^*.h_2$ and $h' = h_1.h_2$, 
where $* \in \{i, o\}$, and $h_1 \vdash M \eqm N.$}\\
\hbox{(flex-rigid)} & h <_f h', \hbox{ if $h = h_1.(\cbf, \cbf)^o.h_2[\cbf/x]$,
$h' = h_1.(x,x)^i.h_2$, $x \not \in \freeFN{h_1}$
and $\cbf \not \in \freeRN{h_1.h_2}$.}
\end{array}
$$
The reflexive-transitive closures of $<_w$, $<_c$ and $<_f$ are
denoted, respectively, by $\sqsubseteq_w$, $\sqsubseteq_c$
and $\sqsubseteq_f.$

If $h \sqsubseteq_f h'$ then $h'$ is obtained from $h$
by substituting certain names, say
$x_1, \ldots, x_n$, 
in $h$ with new rigid names, say, $\cbf_1,\ldots,\cbf_n$,
and changing certain input markings to output. 
In this case, we denote with $\theta_{h,h'}$ the substitution 
$
[\cbf_1/x_1, \ldots, \cbf_n /x_n].
$
\end{definition}

Reading from right-to-left, the above relations read as follows:
The relation $<_w$, called weakening, remove an arbitrary pair from
the bi-trace (hence possibly reducing the knowledge of the observer). 
The relation $<_c$, called contraction, add a redundant pair, i.e., one
which is deducible from the current knowledge, hence adding no extra knowledge.
The relation $<_f$, called flex-rigid, replaces a variable input pair with
a fresh output pair of rigid names. It does not increase the knowledge of the observer, since
the added pair is fresh value, but it does limit the possible respectful substitutions,
since the fresh output pair cannot be substituted (they are rigid names).
Thus, going from right-to-left in the relations, the knowledge of the observer
does not increase. 

\begin{lemma}
\label{lm:order-subst}
Let $h$ and $h'$ be consistent bi-traces and 
let $\vec \theta = (\theta_1,\theta_2)$ be a substitution pair that respects
$h$. For any $t \in \{w, c, f \}$,  if $h \sqsubseteq_t h'$
then $\vec \theta$ respects $h'$ and $h\vec \theta \sqsubseteq_t h'\vec\theta.$
\end{lemma}
\begin{proof}
In all cases, it is obvious that either $h\vec\theta \sqsubseteq_t h'\vec\theta$ holds. 
We therefore need only to show that $\vec\theta$ respects $h'.$
\begin{enumerate}
\item Suppose $h <_w h'$ and $\vec\theta$ respects $h$.
In this case, $h = h_1.h_2$ and $h'=h_1.(M,N)^*.h_2$ for
some $M,$ $N$, $h_1$ and $h_2.$
There are two cases to consider: one in which the weakened
pair $(M,N)$ is an input pair and the other when it is an
output pair.
The latter follows straightforwardly from the definition
of respectful substitutions (which does not impose any requirement
on output pairs) and from the fact that the entailment $\vdash$
is closed under arbitrary extensions of theories (Lemma~\ref{lm:weakening}).
For the former, the proof is by induction on the size of $h_2.$

In the base case, we have $h=h_1$ and $h' = h_1.(M,N)^i$.
We need to show that for every name $x\in \freeFN{M,N}$
we have $h\vec\theta \vdash x\theta_1 \eqm x\theta_2.$
From the definition of $<_w$ we know that all the names
in $M$ and $N$ are also in $h_1.$ And since $\vec\theta$ respects $h_1$,
by Lemma~\ref{lm:respectful-subst}, we have that
$h_1\vec\theta \vdash x\theta_1 \eqm x\theta_2$ for every $x$ in
$\freeFN{h_1}$, hence also for every $x\in \freeFN{M,N}.$
The inductive case follows immediately from the induction hypothesis
and Lemma~\ref{lm:weakening}.

\item Suppose $h <_c h'$ and $\vec\theta$ respects $h.$
There are two cases to consider:
\begin{itemize}
\item $h = h_1.(M,N)^i.h_2$ and $h' = h_1.h_2.$
We show by induction on the length of $h_2$ 
that $\vec\theta$ respects $h'$.
The base case, where $h' = h_1$ and $h = h_1.(M,N)^i$, is obvious, since $\vec\theta$
respects $h$ and therefore it also respects $h'.$
For the inductive cases, the only non-trivial case is when
$h'=h_1.h_2'.(U,V)^i$ and $h = h_1.(M,N)^i.h_2'.(U,V)^i$.
We have to show that $h'\vec\theta \vdash x\theta_1 \eqm x\theta_2$
for every $x\in \freeFN{U,V}.$
Since $\vec\theta$ respects $h$ and $h\vec\theta$ is consistent, we have
$h_1\vec\theta \vdash M\theta_1 \eqm N\theta_2$
and $h\vec\theta \vdash x\theta_1 \eqm x\theta_2.$
Applying Proposition~\ref{prop:cut-elimination} to these two judgments
we therefore obtain 
$
h'\vec\theta\vdash x\theta_1 \eqm x\theta_2
$
as required.

\item $h=h_1.(M,N)^o.h_2$ and $h'=h_1.h_2$. This case is proved by induction
on the length of $h_2$ and Proposition~\ref{prop:cut-elimination}.

\end{itemize}

\item Suppose $h <_f h'$ and $\vec\theta$ respects $h$.
The fact that $\vec\theta$ respects $h'$ can be shown using
the fact that $h'$ and $h$ are essentially equivalent modulo
the injective mapping of names to fresh rigid names: 
for any $M$ and $N$ such that $\cbf \not \in \freeRN{M,N}$,
$h'\vdash M \eqm N$ if and only if $h \vdash M[\cbf/x] \eqm N[\cbf/x].$
This can be shown by a simple induction on the height of the derivation
of the equality.
\end{enumerate}
\qed
\end{proof}

\begin{lemma}
\label{lm:order-ext}
Let $h$ and $h'$ be consistent bi-traces and let $h''$
be a bi-trace such that $h.h''$ is consistent.
Then the following statements hold:
\begin{enumerate}
\item If $h' \sqsubseteq_w h$ and $h' \vdash M \eqm N$
for every $(M,N)^i$ in $h''$, then 
then $h'.h''$ is consistent.

\item If $h' \sqsubseteq_c h$ then $h'.h''$ is consistent.

\item If $h' \sqsubseteq_f h$ then 
$h'.(h''\theta_{h',h})$ is consistent.

\end{enumerate}
\end{lemma}
\begin{proof}
It is sufficient to show the properties hold for the relations
$<_w$, $<_c$ and $<_f.$ In most cases, the proof follows from
inductive arguments, Proposition~\ref{prop:cut-elimination},
Lemma~\ref{lm:weakening} and Lemma~\ref{lm:order-subst}.
\begin{enumerate}
\item Suppose $h' <_w h$.
We show by induction on the size of $h''$ that $h'.h''$ is consistent.
The base case is obvious. 
The inductive cases:
\begin{itemize}
\item $h'' = h_1.(U,V)^i$. We need to show that 
$h'.h'' \vdash U \eqm V$. But this follows from the
assumption that $h'\vdash U \eqm V$. 
\item $h'' = h_1.(U,V)^o$.
We need to show that for every substitution pair $\vec \theta = (\theta_1,\theta_2)$
that respects $h'.h_1$, the theory $\{h'\vec\theta.h''\vec\theta\}$
is consistent. From Lemma~\ref{lm:order-subst}, $\vec\theta$ also respects $h.h_1$,
therefore by the consistency of $h.h''$, 
the theory $\{h\vec\theta.h''\vec\theta\}$ is consistent, which means
that any of its subset is also a consistent theory.
Since $\{h'\vec\theta.h''\vec\theta\} \subseteq \{h\vec\theta.h''\vec\theta\} $
we therefore have that $\{h'\vec\theta.h''\vec\theta\}$ is consistent.
\end{itemize} 

\item Suppose $h' <_c h$. We show that $h'.h''$ is consistent by induction
on the size of $h''.$ We first note that in this case $h$ and
$h'$ are equivalent (as theories), as a consequence of Proposition~\ref{prop:cut-elimination}
and Lemma~\ref{lm:weakening}. 
That is, $h \vdash M \eqm N$ if and only
if $h'\vdash M \eqm N$, for any $M$ and $N.$
The consistency of $h'.h''$ then follows straightforwardly from 
this equivalence, Definition~\ref{def:bi-trace-consistent}, 
Lemma~\ref{lm:order-subst} and induction hypotheses.
 
\item Suppose $h' <_f h$, where $h' = h_1.(\cbf, \cbf)^o.h_2([\cbf/x], [\cbf/x])$
and $h = h_1.(x,x)^i.h_2.$ To show the consistency of $h'.h''[\cbf/x]$ we make use
of the fact that $h' \vdash M[\cbf/x] \eqm N[\cbf/x]$ if and only if
$h \vdash M \eqm N.$ That is, $h$ and $h'$ are indistinguishable as theories. 
The consistency proof then proceeds as in the previous case. 
\end{enumerate}
\qed
\end{proof}

We are now ready to define the up-to techniques. 

\begin{definition}
\label{def:up-to-bisim}
Given a set of consistent traced process pairs ${\cal R}$, define
${\cal R}_t$, for $t \in \{\equiv, w, c, s, i, f, r, p \}$, as the least
relations containing ${\cal R}$ which satisfy the following rules:
\begin{enumerate}
\item up to structural equivalence:
$$
\infer[\equiv]
{h \vdash P ~ {\cal R}_{\equiv} ~ Q}
{P \equiv P', Q \equiv Q' \hbox{ and } h \vdash P'~ {\cal R} ~ Q'}
$$

\item up to weakening:
$$
\infer[w]
{h' \vdash P~{\cal R}_w ~ Q}
{h \vdash P~{\cal R}~Q \hbox{, $h' \sqsubseteq_w h$ and $h'$ is consistent} 
}
$$

\item up to contraction:
$$
\infer[c]
{h' \vdash P~{\cal R}_c ~ Q}
{h \vdash P ~ {\cal R} ~ Q \hbox{, $h' \sqsubseteq_c h$ and $h'$ 
is consistent}}
$$

\item up to substitutions:
$$
\infer[s]
{h\vec \theta \vdash P\theta_1 ~ {\cal R}_s ~ Q\theta_2}
{h\vdash P~{\cal R} ~ Q \hbox{ and $\vec\theta = (\theta_1,\theta_2)$ respects h}}
$$

\item up to injective renaming of rigid names:
$$
\infer[i]
{h(\rho_1,\rho_2) \vdash P\rho_1 ~ {\cal R}_i ~ Q\rho_2}
{h \vdash P ~ {\cal R} ~ Q \hbox{, $\rho_1$ and $\rho_2$ are injective
renaming on rigid names } }
$$

\item up to flex-rigid reversal of names:
$$
\infer[f]
{h' \vdash P\theta_{h',h} ~ {\cal R}_f ~ Q\theta_{h',h}}
{h  \vdash P~{\cal R}~Q, ~ h' \sqsubseteq_f h}
$$

\item up to restriction:
$$
\infer[r]
{h \vdash (\nu \vec x) P ~{\cal R}_r ~ (\nu \vec y)Q}
{
\begin{array}{c}
h \vdash P[\vec \cbf/ \vec x] ~ {\cal R} ~ Q[\vec \dbf/ \vec y], 
\quad \{\vec \cbf \} \cap \freeRN{\pi_1(h),P} = \emptyset, \\
\{\vec \dbf \} \cap \freeRN{\pi_2(h),Q} = \emptyset, \quad \{\vec x, \vec y\} \cap \freeFN{h} = \emptyset
\end{array}
}
$$

\item up to parallel composition:
$$
\infer[p]
{h' \vdash A ~ {\cal R}_p ~ B}
{
\begin{array}{c}
h \vdash P ~ {\cal R} ~ Q, ~ \hbox{ $h'$ is consistent, 
$h' \sqsubseteq_c h$, $\sigma_1 \eqm_{h'} \sigma_2$}, \\ 
\hbox{$\freeFN{R} \subseteq \dom{\sigma_1}$, $\freeRN{R} = \emptyset$,
$A \equiv (P~|~ R\sigma_1)$ and $B \equiv (Q~|~ R\sigma_2).$}
\end{array}
}
$$
\end{enumerate}
{\em Strong open bisimulation up to structural equivalence} is defined
similarly to Definition~\ref{def:open-bisim}, except that we replace
the relation ${\cal R}$ in items (1), (2) and (3) in 
Definition~\ref{def:open-bisim} with ${\cal R}_\equiv$.
Strong open bisimulation up to weakening, contraction, substitutions,
injective renaming, flex-rigid reversal, restrictions and parallel composition
are defined analogously.
\end{definition}

In those rules that concern weakening, contraction and flex-rigid reversal of
names, the observer knowledge in the premise is always equal or greater than
its knowledge in the conclusion. In other words, if the observer cannot
distinguish two processes using its current knowledge, it cannot do so either
in a reduced knowledge. 
In the rule for parallel composition, we allow only processes that can introduce
no extra information to the observer. Notice that in the rule, 
we need to ``contract'' the bi-trace $h$, since we would like to allow $R\sigma_i$ to contain new names not
already in $h$. This does not jeopardize the no-new-knowledge condition, since names
are by default known to observers anyway. 
This flexibility of allowing new names into $R\sigma_i$
will play a (technical) role in showing that the soundness of bisimulation up to parallel
composition.

\begin{lemma}
If ${\cal R}$ is an open bisimulation, then ${\cal R}$ is also
an open bisimulation up to structural equivalence (respectively, 
weakening, contraction, etc.)
\end{lemma}
\begin{proof}
This follows immediately from the fact that 
${\cal R} \subseteq {\cal R}_\equiv$ (respectively, ${\cal R}_w$, etc.).
\qed
\end{proof}

\begin{lemma}
\label{lm:up-to-closure}
Let ${\cal R}$ be a set of consistent traced process pairs. 
Then $({\cal R}_t)_t = {\cal R}_t$, for any $t \in \{\equiv, w, c, s, i, f, r, p \}$.
\end{lemma}

The following lemma states that equivalent substitutions are preserved under
bi-trace extensions. 

\begin{lemma}
\label{lm:subst-equiv-up-to}
Let $h$ and $h'$ be consistent traces such that $h$ is a prefix of $h'$. Let 
$\sigma_1$ and $\sigma_2$ be substitutions such that $\sigma_1 \eqm_h \sigma_2.$
Then $\sigma_1 \eqm_{h'} \sigma_2.$
\end{lemma}

The notions of bisimulation and bisimulation up-to are special cases of the
so called {\em progressions} in \cite{sangiorgi98mscs}. We shall use the techniques
in \cite{sangiorgi98mscs}, adapted to the spi-calculus setting by 
Boreale et.al.\cite{boreale02sjc}, to show that the open bisimulation
relations up-to the closure rules in Definition~\ref{def:up-to-bisim} are sound. 
We first recall some basic notions and results concerning progressions from 
\cite{sangiorgi98mscs}.

\begin{definition}
Given two symmetric and consistent sets of traced process pairs ${\cal R}$
and ${\cal S}$, we say ${\cal R}$ {\em progresses} to ${\cal S}$,
written ${\cal R} \leadsto {\cal S}$, if $h \vdash P ~ {\cal R} ~ Q$
then for all substitution pair $\vec\theta=(\theta_1,\theta_2)$ 
that respects $h$, the following hold:
\begin{enumerate}
\item If $\one{P\theta_1}{\tau}{P'}$ then there exists $Q'$
such that $\one{Q\theta_2}{\tau}{Q'}$ and
$h\vec \theta \vdash P'~{\cal S}~Q'.$
\item If $\one{P\theta_1}{M}{(x)P'}$, where 
$x \not \in \freeFN{h\vec \theta},$
and $\pi_1(h\vec\theta) \vdash M$ 
then there exists $Q'$ such that 
$\one{Q\theta_2}{N}{(x)Q'}$ 
and 
$$
h\vec\theta . (M,N)^i.(x,x)^i \vdash P'~{\cal S}~Q'.
$$
\item If $\one{P\theta_1}{\bar M}{(\nu \vec x)\langle M' \rangle P'},$
and $\pi_1(h\vec\theta) \vdash M$
then there exist $N$, $N'$ and $Q'$ such that $\one{Q\theta_2}{\bar N}{(\nu \vec y)\langle N' \rangle Q'},$
and
$$
h\vec\theta . (M, N)^{i}.(M'[\vec \cbf/\vec x],N'[\vec {\dbf}/\vec y])^o 
\vdash P'[\vec {\cbf}/\vec x] ~{\cal S}~ Q'[\vec {\dbf}/\vec y],
$$
where $\{ \vec \cbf, \vec \dbf \} \cap \freeRN{h\vec\theta,P\theta_1,Q\theta_2} = \emptyset.$ 
\end{enumerate}
\end{definition}

A function ${\cal F}$ on relations is {\em sound} with respect to $\obisim$
if ${\cal R} \leadsto {\cal F}({\cal R})$ implies ${\cal R} \subseteq ~ \obisim.$
${\cal F}$ is {\em respectful} if for every ${\cal R}$ and ${\cal S}$
such that ${\cal R} \subseteq {\cal S}$ and ${\cal R} \leadsto {\cal S}$, 
${\cal F}({\cal R}) \leadsto {\cal F}({\cal S})$ holds. 
We recall some results of \cite{sangiorgi98mscs} regarding respectful functions:
respectful functions are sound, and moreover, compositions of respectful functions
yield respectful functions (hence, sound functions). 
Each rule $t$ in Definition~\ref{def:up-to-bisim} induces a function on relations,
which we denote here with the notation $(.)_t.$ 
We now proceed to showing that the functions induced by the rules in
Definition~\ref{def:up-to-bisim} are sound. We use the notation
$(.)_{t_1\cdots t_n}$ to denote the composition
$(\cdots ((.)_{t_1})_{t_2} \cdots)_{t_n}.$

\begin{lemma}
\label{lm:respectful-function}
The function $(.)_{t}$ for any $t\in \{\equiv, w,c,s,i,f,ri \}$ is respectful.
\end{lemma}
\begin{proof}
Suppose that ${\cal R} \subseteq {\cal S}$. It is easy to see that by
definition, ${\cal R}_t \subseteq {\cal S}_t.$
Moreover, $({\cal R}_t)_t = {\cal R}_t$ for any $t$ and ${\cal R}.$
It remains to show that if ${\cal R} \leadsto {\cal S}$ then
${\cal R}_t \leadsto {\cal S}_t.$
The cases with structural equivalence and injective
renaming follow straightforwardly from the fact that
both preserve one-step transitions. 
The case with substitutions follows straightforwardly from the
fact that compositions of respectful substitutions yield respectful
substitutions (Lemma~\ref{lm:respectful-subst-comp}). 

The cases where $t \in \{w,c,f \}$ are handled uniformly,
following results from Lemma~\ref{lm:order-subst} and Lemma~\ref{lm:order-ext}.
We look at a particular step in the weakening case; the rest can be dealt 
with in a similar fashion.
So let us suppose that $h \vdash P~{\cal R}_w ~ Q$ and
$\vec \theta = (\theta_1,\theta_2)$ respects $h$. The case where
$(h, P, Q) \in {\cal R}$ is trivial, so we look at the other case,
where $h$ is obtained by a weakening step, i.e., $h \sqsubseteq_w h'$
and $h' \vdash P ~ {\cal R} ~ Q$. From Lemma~\ref{lm:order-subst} we know that
$\vec \theta$ respects $h'$ as well.
Now suppose $\one{P\theta_1}{M}{(\nu \vec c) \langle U \rangle P'}$
and $\pi_1(h\vec\theta) \vdash M$ (hence, $\pi_1(h'\vec\theta)\vdash M$).
Since ${\cal R} \leadsto {\cal S}$, 
there exist $N$, $Q'$, $\vec d$ and $V$ such that
$\one{Q\theta_2}{N}{(\nu \vec d)\langle V \rangle Q'}$ and
$$h'\vec\theta.(M,N)^i.(U,V)^o \vdash P' ~ {\cal S} ~ Q'.$$
We need to show that $h\vec\theta.(M,N)^i.(U,V)^o \vdash P' ~ {\cal S}_w ~ Q'.$
We can do this by applying another weakening step to
$h'\vec\theta.(M,N)^i.(U,V)^o \vdash P' ~ {\cal S} ~ Q'.$
To be able do this, we first have to show that the bi-trace
$h\vec\theta.(M,N)^i.(U,V)^o$ is consistent and is a weakening
of $h'\vec\theta.(M,N)^i.(U,V)^o.$ The latter is obvious. For the former,
we note that since $\pi_1(h\vec\theta)\vdash M,$ by the consistency
of $h\vec\theta$, it must be the case that
$h\vec\theta \vdash M \eqm M'$ for a unique $M'$. Now since
$\{h\vec\theta\}$ is a subset of $\{h'\vec\theta\}$, it must be the
case that $h'\vec\theta \vdash M \eqm M'$, and by the consistency of
$h'\vec\theta$, this means that $M' = N.$ In short, we have just
shown that $h\vec\theta \vdash M \eqm N$, therefore we can apply
Lemma~\ref{lm:order-ext} to get the consistency of
$h\vec\theta.(M,N)^i.(U,V)^o$. We can apply the weakening step to
get to
$$h\vec \theta.(M,N)^i.(U,V)^o \vdash P' ~ {\cal S}_w Q'.$$

For the case with $(.)_{ri}$, we first show that
if ${\cal R} \leadsto {\cal S}$ then 
${\cal R}_{r} \leadsto {\cal S}_{ri},$ which is straightforward. The need for
the injective renaming appears when we consider the output transitions, where the choice
of extruded rigid names can vary.
Since we already know that $(.)_i$ is respectful, we have
${\cal R}_{ri} \leadsto {\cal S}_{rii}$. But since ${\cal S}_{rii} = {\cal S}_{ri}$,
we also have ${\cal R}_{ri} \leadsto {\cal S}_{ri}$ as required.
\qed
\end{proof}

In the following, we use the notation $(\vec s, \vec t)^*$, where $*$ is either
an $i$ or an $o$, $\vec s = s_1, \cdots, s_n$, and $\vec t = t_1, \cdots, t_n$, 
to denote the bi-trace $(s_1,t_1)^*.\cdots.(s_n,t_n)^*.$

\begin{proposition}
\label{prop:soundness-bisim-upto}
Let ${\cal R}$ be an open bisimulation up to structural equivalence
(respectively, weakening, contraction, etc.). Then
${\cal R} \subseteq {\cal R}_{\equiv} \subseteq ~ \obisim$
(respectively, ${\cal R} \subseteq {\cal R}_t \subseteq ~ \obisim$,
for $t \in \{w,c, s, i, f, r, p \}$).
\end{proposition}
\begin{proof}
In all cases, ${\cal R} \subseteq {\cal R}_t$ by definition,
so it remains to show ${\cal R}_t \subseteq \  \obisim.$
The case where $t \in \{\equiv, w, c, s, i, f \}$
follows immediately from Lemma~\ref{lm:respectful-function} and the fact that
respectful functions are sound.
For the case with restriction, we first note that since ${\cal R}$
is an open bisimulation up to restriction, we have ${\cal R} \leadsto {\cal R}_r.$
Since ${\cal R} \subseteq {\cal R}_r$, it thus follows from Lemma~\ref{lm:respectful-function}
that ${\cal R}_{ri} \leadsto {\cal R}_{rri}$. Since
${\cal R}_{rri} = {\cal R}_{ri}$, this means that 
${\cal R}_{ri}$ is an open bisimulation and ${\cal R}_{ri} \subseteq \ \obisim.$ 
But since ${\cal R}_r \subseteq {\cal R}_{ri}$, we also have
${\cal R}_r \subseteq ~ \obisim$ as required.

We now look at the case with parallel composition. 
Given that ${\cal R}$ is an open bisimulation up-to parallel composition, we 
show that ${\cal R}_p$ is an open bisimulation up-to substitutions, flex-rigid
reversal, weakening, injective renaming, restriction and structural equivalence. Since all
these up-to bisimulations have been shown to be respectful and sound, 
any of their compositions is also sound, and by showing their inclusion 
of ${\cal R}_p$ we show that ${\cal R}_p$ is included in $\obisim$ as well.

Let us suppose that we are given $h$, $h'$, $P$, $Q$, $R$, $\sigma_1$ and $\sigma_2$
as specified in the rule for ``up to parallel composition'' in 
Definition~\ref{def:up-to-bisim}.
Given $h' \vdash A ~ {\cal R}_p ~ B$ and a subsitution pair $\vec \theta = (\theta_1,\theta_2)$ that
respects $h'$, we examine all the possible transitions from $A$ 
and show that each of these transitions can be matched by $B$ and their continuations 
are in ${\cal R}_{psfw(ri)\equiv}.$
We note that the relation ${\cal R}_{p\vec t}$, where $\vec t$ is a list obtained
from $sfw(ri)\equiv$ by removing one or more function, is contained in 
${\cal R}_{psfw(ri)\equiv}.$
For example, ${\cal R}_{pf(ri)}$ is included in ${\cal R}_{psfw(ri)\equiv}.$
In the following we assume a given substitution pair $\vec\theta = (\theta_1,\theta_2)$
which respects $h'$. Also, we denote with $\rho_1$ and $\rho_2$ the following substitution:
$$
\rho_1 = \rsubst{(\sigma_1 \circ \theta_1)}{\dom{\sigma_1}}
\hbox{ and }
\rho_2 = \rsubst{(\sigma_2 \circ \theta_2)}{\dom{\sigma_2}}.
$$
\begin{enumerate}
\item Suppose $\one{A\theta_1}{\tau}{A'}$ and the transition is driven by $P\theta_1$,
that is, $\one{P\theta_1}{\tau}{P'}$ and $A' \equiv (P' ~|~ R\rho_1)$ (note that
$R\sigma_1\theta_1 = R\rho_1$ by definition).
Since $h \vdash P~{\cal R}~ Q$, ${\cal R}$ is a bisimulation up to parallel composition,
and $\vec \theta$ respects $h$ (Lemma~\ref{lm:order-subst}), we have 
$\one{Q\theta_2}{\tau}{Q'}$ for some $Q'$ such that 
$h\vec\theta \vdash P'~{\cal R}_p Q'.$
By Lemma~\ref{lm:up-to-closure}, $({\cal R}_p)_p = {\cal R}_p$, by Lemma~\ref{lm:comp-subst},
$\rho_1 \eqm_{h\vec\theta} \rho_2$, and since $h'\vec\theta \sqsubseteq_c h\vec\theta$,
it follows from Lemma~\ref{lm:subst-equiv-up-to} that 
$\rho_1 \eqm_{h'\vec\theta} \rho_2.$ We can  therefore apply the up-to-parallel-composition rule
to get 
$$
h'\vec\theta \vdash (P'~|~ R\rho_1) ~ {\cal R}_p ~ (Q' ~|~ R\rho_2)
$$
and 
$$
h'\vec\theta \vdash A' ~ {\cal R}_{p\equiv} ~ B'
$$
for any $B' \equiv (Q' ~ | ~ R\rho_2).$

\item Suppose $\one{A\theta_1}{M}{(x)A'}$, where $\pi_1(h'\vec\theta) \vdash M$, 
and the transition is driven by $P\theta_1$, 
that is, $\one{P\theta_1}{M}{(x)P'}$ and $A' \equiv (P' ~|~ R\rho_1).$ Note that since
we assume processes (and agents) modulo $\alpha$-equivalence, we can assume that $x$
is chosen to be ``fresh'' with respect to the free names in the bi-traces, substitutions 
and processes being considered. We first have to show that $\pi_1(h\vec\theta) \vdash M$
as well; but this is straightforward from the fact that $h'\vec\theta$ is a conservative
extension of $h\vec\theta$.
By similar reasoning to the previous case, we have
$\one{Q\theta_2}{N}{(x)Q'}$ for some $N$ and $Q'$ such that
$
h\vec\theta.(M,N)^i.(x,x)^i \vdash P' ~ {\cal R}_p Q'.
$
Since $h'\vec \vdash M \eqm N$ and $h'\vec\theta \sqsubseteq_c h\vec\theta$,
we have
$$
h'\vec\theta.(M,N)^i.(x,x)^i \sqsubseteq_c h\vec\theta.(M,N)^i.(x,x)^i = h_1
$$
and therefore by Lemma~\ref{lm:subst-equiv-up-to}, we have 
$
\rho_1 \eqm_{h_1} \rho_2.
$
From Lemma~\ref{lm:order-ext}, it follows that $h_1$ is consistent.
This means we can apply the up-to-parallel-composition rule to 
$
h\vec\theta.(M,N)^i.(x,x)^i \vdash P' ~ {\cal R}_p Q'
$
to get 
$
h_1 \vdash (P' ~ | ~ R\rho_1) ~ {\cal R}_p ~ (Q' ~ | ~ R\rho_2)
$
and therefore 
$$
h_1 \vdash A' ~ {\cal R}_{p\equiv} ~ B'
$$
for any $B' \equiv (Q' ~ | ~ R\rho_2).$

\item Suppose $\one{A\theta_1}{\bar M}{(\nu \vec x)\langle M' \rangle A'}$
and the transition is driven by $P\theta_1$, that is 
$\one{P\theta_1}{\bar M}{(\nu\vec x)\langle M' \rangle P'}$
and $A' \equiv (P' ~|~ R\rho_1).$
Then $\one{Q\theta_2}{\bar N}{(\nu \vec y)\langle N' \rangle Q'}$
(therefore, $\one{B}{\bar N}{(\nu\vec y)\langle N' \rangle (Q' ~|~ R\rho_2)}$)
and
$$
h\vec\theta.(M,N)^i.(M'[\vec{\cbf}/\vec x],N'[\vec{\dbf}/\vec y])^o \vdash P'~{\cal R}_p ~Q'.
$$
Let $h_1$ be the bi-trace $h'\vec\theta.(M,N)^i.(M'[\vec{\cbf}/\vec x], N'[\vec{\dbf}/\vec y])^o.$
By Lemma~\ref{lm:order-ext}, $h_1$ is a consistent bi-trace and 
$$
h_1 \sqsubseteq h\vec\theta.(M,N)^i.(M'[\vec{\cbf}/\vec x],N'[\vec{\dbf}/\vec y])^o.
$$
Since $h\vec\theta \sqsubseteq h_1$, it follows from 
Lemma~\ref{lm:subst-equiv-up-to} that $\rho_1 \eqm_{h_1} \rho_2.$ We can now apply the
up-to-parallel-composition rule to get
$$
h_1 \vdash (P' ~|~ R\rho_1) ~ {\cal R}_p ~ {(Q' ~|~ R\rho_2)}
$$
and therefore
$$
h_1 \vdash A' ~ {\cal R}_{p\equiv} ~ B'
$$
for any $B' \equiv (Q' ~|~ R\rho_2).$

\item Suppose $\one{A\theta_1}{\tau}{A'}$ and the transition
is driven by $R\rho_1$, i.e., 
$\one{R\rho_1}{\tau}{R'}$, and $A' \equiv (P\theta_1 ~|~ R')$. 
Then there exists an $U$, $\rho_1'$ and $\rho_2'$  
such that $\rho_1 \preceq \rho_1'$, 
$\rho_2 \preceq \rho_2'$, $\rho_1' \eqm_{h'\vec\theta} \rho_2'$,
$R' = U\rho_1'$ and $\one{R\rho_2}{\tau}{U\rho_2'}.$
Let $U'$ be a renaming of $U$, i.e., $U' = U\rho$ for a renaming
substitution $\rho$, such that $\freeFN{U'} \cap \freeFN{h'} = \emptyset.$ 
Define the substitutions $\delta_1$ and $\delta_2$ as follows:
$$\delta_1 = \rsubst{(\rho^{-1} \circ \rho_1')}{\freeFN{U'}}
\hbox{ and }
\delta_2 = \rsubst{(\rho^{-1} \circ \rho_2')}{\freeFN{U'}}
$$
We note that since $\rho_1' \eqm_{h'\vec\theta} \rho_2'$,
we have $\delta_1 \eqm_{h'\vec\theta} \delta_2.$
Moreover, $U'\delta_1 = U\rho_1'$ and $U'\delta_2 = U\rho_2'$.
Let $\vec x = x_1,\dots,x_n$ be the free names in $U'$. 
Then by the definition of ${\cal R}_p$ we have
$$
h'.(\vec x,\vec x)^i \vdash (P~|~ U') ~ {\cal R}_p (Q ~ | ~ U').
$$
Now let us define $\gamma_1$ and $\gamma_2$ as follows:
$$
\gamma_1 = \theta_1 \circ \delta_1 \hbox{ and } \gamma_2 = \theta_2 \circ \delta_2.
$$
It is easy to see that $\vec \gamma = (\gamma_1,\gamma_2)$ respects
$h'.(\vec x,\vec x)^i$. We can therefore apply the substitution rule to get
$$
h'\vec\gamma.({x_1\gamma_1}, {x_1\gamma_2})^i.\cdots.(x_n\gamma_1,x_n\gamma_2)^i \vdash
(P\gamma_1 ~|~ U'\gamma_1) ~ {\cal R}_{ps} ~ (Q\gamma_2 ~ | ~ U'\gamma_2).
$$

Now since $h'\vec\gamma = h'\vec \theta$ and 
$\freeFN{{x_i\gamma_1}, {x_i\gamma_2}}\subseteq \freeFN{h'\vec\theta}$,
we can apply the weakening rule to get
$$
h'\vec\gamma \vdash
(P\gamma_1 ~|~ U'\gamma_1) ~ {\cal R}_{psw} ~ (Q\gamma_2 ~ | ~ U'\gamma_2)
$$
which is syntactically equivalent to
$$
h'\vec\theta \vdash
(P\theta_1 ~|~ U\rho_1') ~ {\cal R}_{psw} ~ (Q\theta_2 ~ | ~ U\rho_2').
$$
We then apply the congruence rule to get
$$
h'\vec\theta \vdash A' ~ {\cal R}_{psw\equiv} ~ B'
$$
for any $B' \equiv (Q\theta_2 ~|~ U\rho_2').$

\item Suppose $\one{A\theta_1}{M}{(x)A'}$ and the transition is driven
by $R\rho_1$, i.e., $\one{R\rho_1}{M}{(x)R'}$ and $A' \equiv (P\theta_1 ~|~ R')$
(again, here we assume that $x$ is chosen to be sufficiently fresh).
Then there exist $\rho_1'$, $\rho_2'$, $T$ and $U$ such that
$\rho_1 \preceq \rho_1'$, $\rho_2 \preceq \rho_2'$, $\rho_1' \eqm_{h'\vec\theta} \rho_2'$, 
$T\rho_1' = M$ and $U\rho_1' = R'$ and $\one{R\rho_2}{T\rho_2'}{(x)U\rho_2'}.$
In the following discussion, we assume that 
the free names of $T$ and $U$ are distinct from $\freeFN{h'}$, and
that $\dom{\rho_1'} \cap \freeFN{h'} = \emptyset$. This is not a real restriction
since we can use composition with a renaming substitution in the same way 
as in the previous case to avoid name clashes. 
 
Let $\vec y = y_1, \cdots, y_n$ be the free names in $T$ and $U$.
Let $h_1 = h'.(\vec y, \vec y)^i.(T,T)^i.(x,x)^i.$
Since $T$ contains no free rigid names, by Lemma~\ref{lm:equiv-subst} we have
$h' \vdash T \eqm T$, hence $h_1$ is consistent and 
$h_1 \sqsubseteq_c h.$
Therefore  by the definition of ${\cal R}_p$, we have
$$
h'.(\vec y, \vec y)^i.(T,T)^i.(x,x)^i \vdash (P~|~ U) ~ {\cal R}_p ~ (Q~|~ U).
$$
Define $\gamma_1$ and $\gamma_2$ as $\theta_1\circ\rho_1'$ and
$\theta_2\circ \rho_2'$. Clearly $\vec\gamma=(\gamma_1,\gamma_2)$ respects $h_1.$
Therefore, we can apply the substitution rule, with $\vec\gamma$, to get
$$
h'\vec\theta.(y_1\rho_1',y_1\rho_2')^i.\cdots.(y_n\rho_1',y_n\rho_2')^i.
(T\rho_1',T\rho_2')^i.(x,x)^i 
\vdash  (P\theta_1~|~ U\rho_1') ~ {\cal R}_{ps} ~ (Q\theta_2 ~|~ U\rho_2').
$$
Recall that $\rho_1' \eqm_{h'\vec\theta} \rho_2'$, therefore
$\freeFN{y_i\rho_1',y_i\rho_2'} \subseteq \freeFN{h'\vec\theta}$, hence
they can be weakened away:
$$
h'\vec\theta.(T\rho_1',T\rho_2')^i.(x,x)^i 
\vdash  (P\theta_1~|~ U\rho_1') ~ {\cal R}_{psw} ~ (Q\theta_2 ~|~ U\rho_2').
$$
Finally, we apply the structural equivalence rule to get
$$
h'\vec\theta.(T\rho_1',T\rho_2')^i.(x,x)^i 
\vdash  A' ~ {\cal R}_{psw\equiv} ~ B'
$$
where $B' \equiv (Q\theta_2 ~|~ U\rho_2').$

\item Suppose $\one{A\theta_1}{\bar M}{(\nu \vec x)\langle K \rangle A'}$, 
and the transition is driven by $R\rho_1$, 
i.e., $\one{R\rho_1}{M}{(\nu \vec x)\langle K \rangle R'}$ and 
$A' \equiv (\nu \vec x)\langle K \rangle (P\theta_1 ~|~ R'),$
where $\vec x = x_1, \cdots, x_m.$
Then there exist $\rho_1'$, $\rho_2'$, $T$, $L$ and $U$ such that
$\rho_1 \preceq \rho_1'$, $\rho_2 \preceq \rho_2'$, 
$\rho_1' \eqm_{h'\vec\theta} \rho_2',$
$T\rho_1' = M$, $L\rho_1' = K$, $U\rho_1' = R'$ and 
$\one{R\rho_2}{T\rho_2'}{(\nu \vec x)\langle L\rho_2' \rangle U\rho_2'}.$
As in the previous case, we assume, without loss of generality, that
the free names of $T$, $L$, $U$ and the domain of $\rho_1'$
and $\rho_2'$ are all distinct from $\freeFN{h'}.$

Let $\vec y = y_1,\cdots, y_n$ be the free names of $T$ and $L$.
Let $h_1= h'.(\vec y,\vec y)^i.(T,T)^i.(\vec x, \vec x)^i.(L,L)^o.$
Since $T$ and $L$ contain no free rigid names,
we have $h' \vdash T \eqm T$
and $h'.(\vec x, \vec x)^i \vdash L \eqm L.$ Therefore $h_1$ is consistent
and $h_1 \sqsubseteq_c h.$
Let $\gamma_1$ and $\gamma_2$ be defined as 
$\theta_1 \circ \rho_1'$ and $\theta_2\circ \rho_2'$, respectively.
It is easy to verify that $\vec\gamma = (\gamma_1,\gamma_2)$ respects
$h_1$, and $h'\vec\gamma = h'\vec\theta.$
Moreover for every $y_i \in \{y_1,\dots,y_n\}$, 
$\freeFN{y_i\theta_1,y_i\theta_2} \subseteq \freeFN{h'\vec\theta}.$

We can then apply the following series of rules:
$$
\begin{array}{c}
h \vdash P ~ {\cal R} ~ Q \\
\Downarrow p \\
h'.(\vec y, \vec y)^i.(T,T)^i.(\vec x, \vec x)^i.(L,L)^o 
\vdash (P~|~U)  ~ {\cal R}_p ~ (Q~|~ U) \\
\Downarrow s \\
h'\vec\theta.(y_1\rho_1',y_1\rho_2')^i.\cdots.(y_n\rho_1',y_n\rho_2')^i.
(T\rho_1', T\rho_2')^i.(\vec x, \vec x)^i.(L\rho_1',L\rho_2')^o
\vdash (P\theta_1 ~|~ U\rho_1') ~ {\cal R}_{ps} ~ (Q\theta_2 ~ | ~ U\rho_2')\\
\Downarrow f \\
h'\vec\theta.(y_1\rho_1',y_1\rho_2')^i.\cdots.(y_n\rho_1',y_n\rho_2')^i.
(T\rho_1',T\rho_2')^i.(\vec {\cbf}, \vec {\cbf})^o.
(L\rho_1'[\vec {\cbf}/\vec x],L\rho_2'[\vec {\cbf}/\vec x])^o \\
\vdash (P\theta_1 ~|~ U\rho_1'[\vec {\cbf}/\vec x]) ~ {\cal R}_{psf} ~ 
(Q\theta_2 ~ | ~ U\rho_2'[\vec {\cbf}/\vec x]) \\
\Downarrow w\\
h'\vec\theta.(T\rho_1',T\rho_2')^i.(L\rho_1'[\vec {\cbf}/\vec x],L\rho_2'[\vec {\cbf}/\vec x])^o
\vdash (P\theta_1 ~|~ U\rho_1'[\vec {\cbf}/\vec x]) ~ {\cal R}_{psfw} ~ 
(Q\theta_2 ~ | ~ U\rho_2'[\vec {\cbf}/\vec x]) \\
\Downarrow \equiv \\
h'\vec\theta.(T\rho_1',T\rho_2')^i.(L\rho_1'[\vec {\cbf}/\vec x],L\rho_2'[\vec {\cbf}/\vec x])^o
\vdash A'[\vec {\cbf}/\vec x] ~ {\cal R}_{psfw\equiv} ~ 
B'[\vec {\cbf}/\vec x] 
\end{array}
$$ 
where $B' \equiv (Q\theta_2 ~ | ~ U\rho_2'[\vec{\cbf}/\vec x]).$

\item Suppose that $\one{A\theta_1}{\tau}{A'}$ and the transition is driven by
an output action by $P\theta_1$ and an input action by $R\rho_1$. That is, 
$\one{P\theta_1}{\bar M}{(\nu \vec y)\langle M_1 \rangle P'}$
and $\one{R\rho_1}{M}{(x)R'}$ and
$A'\equiv (\nu \vec y)(P' ~|~ R'[M_1/x]).$
Then we have 
\begin{itemize}
\item $\one{Q\theta_2}{\bar N}{(\nu \vec z)\langle N_1 \rangle Q'}$
and 
$h'\vec\theta.(M,N)^i.(M_1[\vec {\cbf}/\vec x], N_1[\vec {\dbf}/\vec z])^o
\vdash P'~{\cal R}_p ~ Q'$, and
\item there exist $\rho_1'$, $\rho_2'$, $T$ and $U$ such that 
$\rho_1 \preceq \rho_1'$, $\rho_2 \preceq \rho_2'$,
$\rho_1' \eqm_{h'\vec\theta} \rho_2'$, $T\rho_1' = M$ and $U\rho_1' = R'$ and 
$\one{R\rho_2}{T\rho_2'}{(x)U\rho_2'}.$
\end{itemize}
By Lemma~\ref{lm:equiv-subst}, we know that $h'\vec\theta \vdash T\rho_1' \eqm T\rho_2'.$
Since $h'\vec\theta$ is consistent, and $T\rho_1' = M$, it must be the case
that $T\rho_2' = N.$
Let $h_1 = h\vec\theta.(M,N)^i.(M_1[\vec {\cbf}/\vec y], N_1[\vec{\dbf}/\vec z])^o$
and let $h_2 = h'\vec\theta.(M,N)^i.(M_1[\vec {\cbf}/\vec y], N_1[\vec{\dbf}/\vec z])^o.$
Obviously, $h_2 \sqsubseteq_c h_1$ and since $h_1$ is consistent, by Lemma~\ref{lm:order-ext}, 
we have that $h_2$ is also consistent.
Now define $\sigma_1'$ and $\sigma_2'$ as follows
$$
\sigma_1' = \rho_1' \cup \{x \mapsto M_1[\vec {\cbf}/\vec y] \}
\qquad
\hbox{ and }
\sigma_2' = \rho_2' \cup \{x \mapsto N_1[\vec {\dbf}/\vec z] \}.
$$
It is easy to see that $\sigma_1' \eqm_{h_2} \sigma_2'.$
We can now apply the following series of rules
$$
\begin{array}{c}
h\vec\theta.(M,N)^i.(M_1[\vec {\cbf}/\vec y], N_1[\vec{\dbf}/\vec z])^o
\vdash P' ~{\cal R}_p Q' \\
\Downarrow p \\
h'\vec\theta.(M,N)^i.(M_1[\vec {\cbf}/\vec y], N_1[\vec{\dbf}/\vec z])^o 
\vdash (P' ~ | ~ U\sigma_1') ~{\cal R}_p ~ (Q' ~|~ U\sigma_2') \\
\Downarrow w\\
h'\vec\theta \vdash (P' ~ | ~ U\sigma_1') ~{\cal R}_{pw} ~ (Q' ~|~ U\sigma_2') \\
\Downarrow ri\\
h'\vec\theta \vdash (\nu \vec y)(P' ~ | ~ U\rho_1'[M_1/x]) ~{\cal R}_{pw(ri)} ~ 
(\nu \vec z)(Q' ~|~ U\rho_2'[N_1/x]) \\
\Downarrow \equiv\\
h'\vec\theta \vdash A' ~ {\cal R}_{pw(ri)\equiv} ~ B'
\end{array}
$$
where $B' \equiv (\nu \vec z)(Q' ~|~ U\rho_2'[N_1/x]).$

\item Suppose $\one{A\theta_1}{\tau}{A'}$ and the transition is driven by
an input by $P\theta_1$ and an output by $R\rho_1.$
That is, $\one{P\theta_1}{M}{(x)P'}$ and $\one{R\rho_1}{\bar M}{(\nu \vec y)\langle M_1 \rangle R'}$
and $A' \equiv (\nu \vec y)(P'[M_1/x] ~|~ R').$
Then we have
\begin{itemize}
\item $\one{Q\theta_2}{N}{(x)Q'}$ and 
$h\vec\theta.(M,N)^i.(x,x)^i \vdash P'~{\cal R}_p ~ Q'$, and
\item there exist $\rho_1'$, $\rho_2'$,  $T$, $K$ and $U$ such that 
$\rho_1 \preceq \rho_1'$, $\rho_2 \preceq \rho_2'$, 
$\rho_1' \eqm_{h'\vec\theta} \rho_2'$,
$T\rho_1' = M$, 
$K\rho_1' = M_1$, $U\rho_1' = R'$ (we can assume w.l.o.g. that $\vec y$
are fresh w.r.t. $\rho_1'$ and $\rho_2'$) and 
$\one{R\rho_2}{T\rho_2'}{(\nu \vec y)\langle K\rho_2' \rangle U\rho_2'}.$
\end{itemize}
Using a similar argument as in the previous case, we can show
that $T\rho_2' = N.$
Let us now construct a bi-trace as follows:
$$
h_1 = h'\vec\theta.(M,N)^i.(\vec y,\vec y)^i.(K\rho_1',K\rho_2')^o.(x,x)^i.
$$
It is straightforward to show that 
$$
h_1 \sqsubseteq_c h\vec\theta.(M,N)^i.(x,x)^i,
$$
that $h_1$ is consistent (it is sufficient to show
that $h'\vec\theta \vdash K\rho_1' \eqm K\rho_2'$, using Lemma~\ref{lm:equiv-subst})
and that $\rho_1' \eqm_{h_1} \rho_2'.$ 
In the following, we use the following denotations for some terms:
\begin{itemize}
\item $M_1 = K\rho_1'$, $N_1 = K\rho_2'$,
\item $M_2 = M_1[\vec {\cbf}/\vec y]$, $N_2 = N_1[\vec {\cbf}/\vec y]$,
\end{itemize}
where $\{\vec \cbf\} \cap \freeRN{h'\vec\theta} = \emptyset.$
We can now apply the following up-to rules:
$$
\begin{array}{c}
h\vec\theta.(M,N)^i.(x,x)^i \vdash P' ~{\cal R}_p ~ Q' \\
\Downarrow p \\
h'\vec\theta.(M,N)^i.(\vec y,\vec y)^i.(M_1,N_1)^o.(x,x)^i
\vdash (P' ~ | ~ U\rho_1') ~ {\cal R}_p ~ (Q' ~|~ U\rho_2') \\
\Downarrow s \\
h'\vec\theta.(M,N)^i.(\vec y,\vec y)^i.(M_1,N_1)^o.(M_1,N_1)^i
\vdash (P'[M_1/x] ~ | ~ U\rho_1') ~ {\cal R}_{ps} ~ (Q'[N_1/x] ~ | ~ U\rho_2')\\
\Downarrow f\\
h'\vec\theta.(M,N)^i.(\vec {\cbf},\vec {\cbf})^o.(M_2,N_2)^o.(M_2,N_2)^i
\vdash (P'[M_2/x] ~ | ~ U\rho_1'[\vec {\cbf}/\vec y]) ~ 
{\cal R}_{psf} ~ (Q'[N_2/x] ~ | ~ U\rho_2'[\vec {\cbf}/\vec y])\\
\Downarrow w\\
h'\vec\theta \vdash (P'[M_2/x] ~ | ~ U\rho_1'[\vec {\cbf}/\vec y]) ~ 
{\cal R}_{psfw} ~ (Q'[N_2/x] ~ | ~ U\rho_2'[\vec {\cbf}/\vec y])\\
\Downarrow ri\\
h'\vec\theta \vdash (\nu \vec y)(P'[M_1/x] ~ | ~ U\rho_1') ~ {\cal R}_{psfw(ri)} ~ 
(\nu \vec y)(Q'[N_1/x] ~|~ U\rho_2')\\
\Downarrow \equiv\\
h'\vec\theta \vdash A' ~ {\cal R}_{psfw(ri)\equiv} ~ B'
\end{array}
$$
where $B' \equiv (\nu \vec y)(Q'[N_1/x] ~ | ~ U\rho_2').$
\end{enumerate}
\qed
\end{proof}

\begin{corollary}
For every $t\in \{w,c, s, i, f, r, p \}$, $(\obisim)_t = \ \obisim.$
\end{corollary}

\section{Soundness of open bisimilarity}
\label{sec:sound}

We now show that open bisimilarity is sound with respect to testing equivalence.

\begin{theorem}
\label{thm:soundness}
If $P \sim_o Q$ then $P \sim Q$.
\end{theorem}
\begin{proof}
Suppose $P \sim_o Q$. Note that by Definition~\ref{def:open-bisim}, $P$ and $Q$ are
pure processes. Let $R$ be a pure process. 
We have to show that the transitions of $(P~|~R)$ can be matched by $(Q~|~R)$
and vice versa. We show here the first case, the other case  can be proved using
a symmetric argument.

Suppose
$$
P~|~R \stackrel{\tau}{\longrightarrow} P_1 \stackrel{\tau}{\longrightarrow} \cdots \stackrel{\tau}{\longrightarrow}
P_n \stackrel{\beta}{\longrightarrow} A,
$$
for some $P_1, \ldots, P_n, \beta$ and $A.$ We show that this sequence of transitions
can be matched by $Q.$ Note that since both $P$ and $R$ are pure processes,
every $P_i$ is also a pure process. 
Since $P\sim_o Q$, we have $h \vdash P \obisim Q$ for some universal bi-trace $h$. 
Since $\obisim$ is closed under bi-trace contraction, we can assume without loss of 
generality that $h$ contains all the free names of $P$,$Q$ and $R.$
By Proposition~\ref{prop:soundness-bisim-upto}, we have  
$h \vdash (P~|~R) \obisim (Q ~ | R)$, which means that, 
by Definition~\ref{def:open-bisim}, there are $Q_1, \ldots, Q_n$ such that
$$
Q~|~R \stackrel{\tau}{\longrightarrow} Q_1 \stackrel{\tau}{\longrightarrow} \cdots \stackrel{\tau}{\longrightarrow} Q_n 
$$
and $h \vdash P_i \obisim Q_i$ for each $i \in \{1,\ldots, n\}$. 
In particular, $h \vdash P_n \obisim Q_n$, therefore we have
$$
Q_n \stackrel{\beta'}{\longrightarrow} B
$$
for some $B$ and $\beta'$ such that $h \vdash \beta \eqm \beta'.$
But since $\beta$ contains no rigid names, by Lemma~\ref{lm:open-message}, it must be the case
that $\beta' = \beta.$ We therefore have 
$$
Q~|~R \stackrel{\tau}{\longrightarrow} Q_1 \stackrel{\tau}{\longrightarrow} \cdots \stackrel{\tau}{\longrightarrow}
Q_n \stackrel{\beta}{\longrightarrow} B.
$$
\qed
\end{proof}

\section{An example}
\label{sec:ex}


This example demonstrates the use of the up-to techniques in proving
bisimilarity. This example is adapted from a similar one in \cite{borgstrom04concur}.
Let $P$ and $Q$ be the following processes:
$$
P = \abf(x).(\nu k)\spout \abf {\enc x k} {(\nu m)\spout \abf {\enc m {\enc \abf k}} {\spout m \abf 0}}
$$
$$
Q = \abf(x).(\nu k)\spout \abf {\enc x k} {(\nu m)\spout \abf {\enc m {\enc \abf k}} {[x=\abf]\spout m \abf 0}}
$$
Let ${\cal R}$ be the least set such that:
$$
\begin{array}{l}
 (\abf,\abf)^o  \vdash  P ~ {\cal R} ~ Q, \quad (\abf,\abf)^o.(x,x)^i \vdash   P_1 ~ {\cal R} ~ Q_1, \\
 (\abf,\abf)^o.(x,x)^i.(\enc x \kbf, \enc x \kbf)^o \vdash P_2 ~ {\cal R} ~ Q_2,\\
 (\abf,\abf)^o.(x,x)^i.(\enc x \kbf, \enc x \kbf)^o.(\enc \mbf {\enc \abf \kbf}, \enc \mbf {\enc \abf \kbf})^o
 \vdash P_3 ~ {\cal R} ~ Q_3,\\
 (\abf,\abf)^o.(\abf,\abf)^i.(\enc \abf \kbf, \enc \abf \kbf)^o.(\enc \mbf {\enc \abf \kbf}, 
\enc \mbf {\enc \abf \kbf})^o.(\mbf, \mbf)^i.(\abf,\abf)^o  \vdash 
  0 ~ {\cal R} ~ 0,
\end{array}
$$
where 
$$
\begin{array}{c}
P_1 = (\nu k)\spout \abf {\enc x \kbf} {(\nu m)\spout \abf {\enc m {\enc \abf k}} {\spout m \abf 0}}, \\
Q_1 = (\nu k)\spout \abf {\enc x k} {(\nu m)\spout \abf {\enc m {\enc \abf k}} {[x=\abf]\spout m \abf 0}}, \\
P_2 = (\nu m)\spout \abf {\enc m {\enc \abf \kbf}} {\spout m \abf 0}, \quad
Q_2 = (\nu m)\spout \abf {\enc m {\enc \abf \kbf}} {[x=\abf]\spout m \abf 0}, \\
P_3 = \spout \mbf \abf 0, ~ Q_3  = [x = \abf]\spout \mbf \abf 0.
\end{array}
$$
Let ${\cal R}'$ be the symmetric closure of ${\cal R}$. Then it is easy to
see that ${\cal R}'$ is an open bisimulation up-to contraction and substitutions. 
For instance, consider the traced process pair
$
h \vdash  \spout \mbf \abf 0 ~ {\cal R}' ~ [x=\abf] \spout \mbf \abf 0 \\
$
where $h=(\abf,\abf)^o.(x,x)^i.(\enc x \kbf, \enc x \kbf)^o.(\enc \mbf {\enc \abf \kbf},
 \enc \mbf {\enc \abf \kbf})^o.$
Let $\vec\theta = (\theta_1,\theta_2)$ be an $h$-respectful substitution.
Since $x$ is the only name in $h$, we have
$$
h\vec\theta = (\abf,\abf)^o.(s,t)^i.(\enc s \kbf, \enc t \kbf)^o.(\enc \mbf {\enc \abf \kbf}, 
\enc \mbf {\enc \abf \kbf})^o,
$$
where $s=x\theta_1$ and $t=x\theta_2.$
We have to check that every detectable action from $\spout \mbf \abf 0$
can be matched by $[t=\abf]\spout \mbf \abf 0.$ 
If $t \not = \abf$, then $s \not = \abf$ (by the consistency of $h\vec\theta$),
therefore, $\pi_1(h\vec\theta) \not \vdash \mbf$, i.e., the action
$\mbf$ is not detected by the environment, so this case is trivial. 
If $t = \abf$, then $s = \abf$ and $h\vec\theta \vdash \mbf \eqm \mbf$, so both $P_3\theta_1$ and $Q_3\theta_2$
can make a transition on channel $\mbf$. Their continuation is
the traced process pair
$$
(\abf,\abf)^o.(\abf,\abf)^i.(\enc \abf \kbf, \enc \abf \kbf)^o.(\enc \mbf {\enc \abf \kbf}, 
\enc  \mbf {\enc \abf \kbf})^o.(\mbf, \mbf)^i.(\abf,\abf)^o
\vdash 0 ~ {\cal R}' ~ 0
$$
which is in the set ${\cal R}'$, hence also in ${{\cal R}'}_{cs}$ (up-to contraction
and substitution on ${\cal R}'$). Therefore by Proposition~\ref{prop:soundness-bisim-upto}, 
$(\abf,\abf)^o \vdash P ~ \obisim ~ Q.$

\section{Congruence results for open bisimilarity}
\label{sec:congr}

In this section we show that the relation $\sim_o$ on pure
processes is an equality relation (reflexive, 
symmetric, transitive) and is closed under arbitrary
pure process contexts.
We need some preliminary lemmas to show that $\sim_o$ is an
equivalence relation. Most of these lemmas concern properties
of {\em reflexive observer theories}, i.e., theories in which their
first and second projections are equal sets.

\begin{lemma}
\label{lm:open-message}
Let $M$ be a pure message. Then $\Gamma \vdash M \eqm M$
for any theory $\Gamma.$
\end{lemma}

\begin{lemma}
\label{lm:reflexive-theory}
Let $\Gamma$ be a theory such that $\pi_1(\Gamma) = \pi_2(\Gamma).$ 
If $\Gamma \vdash M \eqm N$, then $M = N.$
\end{lemma}
\begin{proof}
By simple induction on the height of the derivation
of $\Gamma \seqsym M \eqm N.$ \qed
\end{proof}

\begin{lemma}
\label{lm:reflexive-theory-consistent}
Let $\Gamma$ be a theory such that $\pi_1(\Gamma) = \pi_2(\Gamma)$.
Then $\Gamma$ is a consistent theory.
\end{lemma}
\begin{proof}
We show that $\Gamma$ satisfies the list of properties specified
in Definition~\ref{def:theory-consistency}. The first and the third
properties follow immediately from Lemma~\ref{lm:reflexive-theory}.
For the second property, we need to show that whenever 
$\Gamma \vdash \enc M N \eqm \enc M N$, then 
$\pi_1(\Gamma) \vdash N$ (or $\pi_2(\Gamma)\vdash N$) 
implies $\Gamma\vdash N \eqm N.$ This can be proved straightforwardly
by induction on the length of derivations, that is, we simply mimic the
rules applied in $\pi_i(\Gamma) \vdash N$ to prove
$\Gamma \vdash N \eqm N.$
\qed
\end{proof}

\begin{lemma}
\label{lm:reflexive-bi-trace}
Let $h$ be a consistent bi-trace such that $\pi_1(h) = \pi_2(h)$.
If $\vec \theta = (\theta_1,\theta_2)$ respects $h$,
then $\pi_1(h\vec\theta) = \pi_2(h\vec\theta)$
and for every $x\in \freeFN{h}$, $x\theta_1 = x\theta_2.$
\end{lemma}
\begin{proof}
By induction on the size of $h$. The non-trivial case is
when $h = h'.(M,M)^i$. By the induction hypothesis,
we have that $\pi_1(h'\vec \theta) = \pi_2(h'\vec \theta)$,
therefore by Lemma~\ref{lm:reflexive-theory},
$M\theta_1 = M\theta_2.$
Moreover, since $\vec\theta$ respects $h$, it is the case
that $h'\vec\theta \vdash x\theta_1 \eqm x\theta_2$,
and again by Lemma~\ref{lm:reflexive-theory},
$x\theta_1 = x\theta_2.$
\qed
\end{proof}

\begin{lemma}
\label{lm:output-extension}
Let $h=h'.(M,M)^o$ be a bi-trace such that $h'$ is consistent,
$\pi_1(h') = \pi_2(h')$ and $\freeFN{M}\subseteq \freeFN{h'}.$
Then $h$ is a consistent bi-trace.
\end{lemma}
\begin{proof}
We have to show that for every $h'$-respectful substitution
pair $\vec \theta = (\theta_1,\theta_2)$,
$\{h\vec\theta\}$ is a consistent theory.
From Lemma~\ref{lm:reflexive-bi-trace}, it follows that
$\pi_1(h'\vec \theta) = \pi_2(h'\vec \theta)$.
And since $\freeFN{M} \subseteq \freeFN{h'}$, 
we have $M\theta_1 = M\theta_2$ and 
$\pi_1(h\vec \theta) = \pi_2(h\vec \theta)$.
Therefore by Lemma~\ref{lm:reflexive-theory-consistent}, 
$\{h\vec \theta\}$ is a consistent theory. Thus, 
$h$ is a consistent bi-trace. 
\qed
\end{proof}

\begin{lemma}
\label{lm:open-bisim-reflexive}
The set 
$$
{\cal R} = \{ (h,P,P) \mid \hbox{$(h,P,P)$ is a traced process pair,
$h$ is consistent and $\pi_1(h) = \pi_2(h)$} \}
$$
is an open bisimulation.
\end{lemma}
\begin{proof}
${\cal R}$ is obviously symmetric and consistent.
It remains to show that it is closed under one-step transitions.
Suppose $h \vdash P ~ {\cal R} ~ P$ and 
$\vec \theta = (\theta_1,\theta_2)$ respects $h$.
Note that $P\theta_1 = P\theta_2$ since $\theta_1$ and $\theta_2$
coincide on the domain $\freeFN{h}$ by Lemma~\ref{lm:reflexive-bi-trace} 
(recall that the free names of $P$ are among the free names in $h$).
\begin{enumerate}
\item Suppose $\one{P \theta_1}{\tau}{P'}$. Since $P\theta_1 = P\theta_2$,
we have $\one{P\theta_2}{\tau}{P'}$, and since $h\vec\theta$
is consistent, we have $h\vec\theta \vdash P'~{\cal R}~P'.$

\item Suppose $\one{P\theta_1}{M}{(x)P'}$, 
$x \not \in \freeFN{h\vec\theta}$, and $\pi_1(h\vec\theta) \vdash M$. 
Then $\one{P\theta_2}{M}{(x)P'}$, and  since 
$h\vec\theta$ is consistent, by Lemma~\ref{lm:unique-projection}, we have
$h\vec\theta \vdash M \eqm N$ for some $N$. By Lemma~\ref{lm:reflexive-theory},
we have $N = M.$ This, together with the fact that
$h\vec\theta.(M,M)^i \vdash x \eqm x$, entail
that $h\vec\theta.(M, M)^i.(x,x)^i$ is consistent and  therefore
$$
h\vec\theta.(M,M)^i.(x,x)^i \vdash P'~{\cal R}~P'.
$$

\item Suppose $\one{P\theta_1}{\bar M}{(\nu \vec x)\langle N \rangle P'},$
and $\{\vec \cbf \} \cap \freeRN{h\vec\theta,P\theta_1,Q\theta_2} = \emptyset$,
and $\pi_1(h\vec\theta) \vdash M$. Then 
$\one{P\theta_2}{\bar M}{(\nu \vec x)\langle N \rangle P'}$ and
following the same argument as in the previous case,
we show that $h\vec \theta.(M,M)^i$ is consistent.
From Lemma~\ref{lm:output-extension} it follows that 
$h\vec\theta.(M,M)^i.(N[\vec \cbf/\vec x],N[\vec \cbf/\vec x])^o$ is also consistent, therefore
$$
h\vec\theta.(M,M)^i.(N[\vec \cbf/\vec x],N[\vec \cbf/\vec x])^o 
\vdash P'[\vec \cbf/\vec x] ~{\cal R}~ P'[\vec \cbf/\vec x].
$$
\end{enumerate}
\qed
\end{proof}

\begin{definition}
Given two sets of traced process pairs ${\cal R}_1$ and
${\cal R}_2$, their composition is defined as follows:
$$
{\cal R}_1 \circ {\cal R}_2 =
\{ (h_1\circ h_2, P, R) \mid
\hbox{$h_1 \vdash P~{\cal R}~Q$, $h_2 \vdash Q~{\cal R}_2 ~ R$
and $h_1$ is left-composable with $h_2$} 
\}.
$$
\end{definition}

\begin{lemma}
\label{lm:open-bisim-comp}
If ${\cal R}_1$ and ${\cal R}_2$ are open bisimulations
then ${\cal R}_1 \circ {\cal R}_2$ is also an open bisimulation.
\end{lemma}
\begin{proof}
The symmetry of ${\cal R}_1 \circ {\cal R}_2$ follows from the
symmetry of ${\cal R}_1$ and ${\cal R}_2$
and its consistency follows from the fact that compositions
of consistent bi-traces yield consistent bi-traces 
(Lemma~\ref{lm:trace-comp-consistent}).
It remains to show that ${\cal R}_1 \circ {\cal R}_2$ is closed
under one-step transitions. In the following ${\cal R}$ denotes
the set ${\cal R}_1 \circ {\cal R}_2$.
Suppose $h_1 \circ h_2 \vdash P ~ {\cal R} ~ R$ 
and $\vec \theta = (\theta_1, \theta_2)$ respects $h_1 \circ h_2$. 
From the definition of ${\cal R}$ we have that
$h_1 \vdash P ~ {\cal R}_1 ~ Q$ and
$h_2 \vdash Q ~ {\cal R}_2 ~ R$ for some $Q$.
It follows from Lemma~\ref{lm:separating-subst} that there exists a substitution $\rho$
such that $(\theta_1,\rho)$ respects $h_1$ and
$(\rho,\theta_2)$ respects $h_2.$

\begin{enumerate}
\item Suppose $\one{P\theta_1}{\tau}{P'}$. 
Then $\one{Q\rho}{\tau}{Q'}$ and
$\one{R\theta_2}{\tau}{R'}$ for some $Q'$ and $R'$
such that $h_1(\theta_1,\rho) \vdash P'~{\cal R}_1 ~ Q'$
and $h_2(\rho,\theta_2) \vdash Q'~{\cal R}_2 ~ R'$.
Therefore $(h_1 \circ h_2)\vec \theta \vdash P' ~ {\cal R} ~ R'.$

\item Suppose $\one{P\theta_1}{M}{(x)P'}$, where $x\not \in \freeFN{h\vec\theta}$
and $\pi_1((h_1\circ h_2)\vec \theta) \vdash M.$
Then $\one{Q\rho}{N}{(x)Q'}$ and $\one{R\theta_2}{U}{(x)R'}$ for some $N$, $U$, 
$Q'$ and $R'$ such that
\begin{itemize}
\item $h_1(\theta_1,\rho).(M,N)^i.(x,x)^i \vdash P'~{\cal R}_1 ~ Q',$ and
\item $h_2(\rho, \theta_2).(N,U)^i.(x,x)^i \vdash Q' ~ {\cal R}_2 ~ R'$.
\end{itemize}
Therefore $(h_1\circ h_2)\vec \theta.(M,U)^i.(x,x)^i \vdash P'~ {\cal R} ~ R'.$

\item Suppose $\one{P\theta_1}{\bar M}{(\nu \vec x)\langle M' \rangle P'}$
for some $M$, $M'$ and $P'.$
Then $\one{Q\rho}{\bar N}{(\nu \vec y)\langle N' \rangle Q'}$
and $\one{R\theta_2}{\bar U}{(\nu \vec z)\langle U' \rangle R'}$ 
for some $Q'$, $R'$, $N$,$U$, $N'$ and $U'$ such that 
\begin{itemize}
\item $h_1(\theta_1,\rho).(M,N)^i.(M'[\vec \cbf/\vec x],N'[\vec \dbf/\vec y])^o 
\vdash P'[\vec \cbf/\vec x] ~ {\cal R}_1  ~ Q'[\vec \dbf/\vec y]$, and
\item $h_2(\rho,\theta_2).(N,U)^i.(N'[\vec \dbf/\vec y],U'[\vec \ebf/\vec y])^o 
\vdash Q'[\vec \dbf/\vec y] ~ {\cal R}_2 ~ R'[\vec \ebf/\vec y],$
\end{itemize}
where $\vec \cbf$, $\vec \dbf$ and $\vec \ebf$ satisfy the freshness
condition in Definition~\ref{def:open-bisim}.
Therefore 
$$
(h_1\circ h_2)\vec\theta.(M,U)^i.(M'[\vec \cbf/\vec x],U'[\vec \ebf/\vec z])^o 
\vdash P'[\vec \cbf/\vec x]~{\cal R}~R'[\vec \ebf/\vec z].
$$
\end{enumerate}
\qed
\end{proof}

\begin{theorem}
\label{thm:equiv-relation}
The relation $\sim_o$ is an equivalence relation on pure processes. 
\end{theorem}
\begin{proof}
The symmetry of $\sim_o$ follows from the symmetry of $\obisim$.
For the reflexivity, from Lemma~\ref{lm:open-bisim-reflexive} we know that there is
a bisimulation ${\cal R}$ that contains $(h, R, R)$ for
any pure process $R$ and any universal trace $h$ such that $\freeFN{R} \subseteq \freeFN{h}$. 
Therefore ${\cal R} \subseteq \obisim$ and
$R \sim_o R$ for all pure process $R$.
For transitivity, from Lemma~\ref{lm:open-bisim-comp} we know that $(\obisim) \circ (\obisim)$
is an open bisimulation, hence $(\obisim) \circ (\obisim) \subseteq ~ \obisim$ (because
$\obisim$ is the largest open bisimulation).
Now suppose $P \sim_o Q$ and $Q \sim_o R$. This means that for some $h_1$
and $h_2$, $(h_1,P,Q) \in \obisim$ and $(h_2,Q,R) \in \obisim$.
Using Proposition~\ref{prop:soundness-bisim-upto}, we can introduce arbitrary
pairs of input names to a traced process pair while still preserving their
bisimilarity. It thus follows that there is an $h$ such that
$\freeFN{h_1,h_2} \subseteq \freeFN{h}$, 
$(h,P,Q) \in \obisim$ and $(h,Q,R) \in \obisim$. 
Therefore, by Lemma~\ref{lm:open-bisim-comp}, $(h,P,R) \in \obisim$, hence $P \sim_o R.$
\qed
\end{proof}

Having established that $\sim_o$ is indeed an equivalence relation on pure
processes, we proceed to showing that it is also a congruence, for {\em finite} pure processes.

\begin{lemma}
\label{lm:clo-input}
$h.(x,x)^i \vdash P ~ \obisim ~ Q$ if and only if
$h \vdash M(x).P ~ \obisim ~ N(x).Q$ where $h \vdash M \eqm N$ and $x\not\in \freeFN{h}.$
\end{lemma}
\begin{proof}
Suppose $h.(x,x)^i \vdash P ~ \obisim ~ Q$. Then there exists an
open bisimulation ${\cal R}$ such that $h.(x,x)^i \vdash P ~ {\cal R} ~ Q.$
Define the relation ${\cal R}_i$ as follows:
$$
{\cal R}_i = \{(h, M(x).P, N(x).Q) \mid h.(x,x)^i \vdash P ~ {\cal R} ~ Q \hbox{ and }
h\vdash M \eqm N \}.
$$
It is easy to show that ${\cal R}_i$ is an open bisimulation, therefore,
$h\vdash M(x).P ~ \obisim ~ N(x).Q$ for any $h \vdash M \eqm N.$

Conversely, suppose that $h\vdash M(x).P ~ {\cal R} ~ N(x).Q$ and
$h\vdash M \eqm N$, for some open bisimulation ${\cal R} \subseteq ~ \obisim.$
Since the empty substitution pair $(\epsilon,\epsilon)$ respects $h$
and since $\one{M(x).P}{M}{(x)P}$ and $\one{N(x).Q}{N}{(x)Q}$,
we obviously have $h.(M,N)^i.(x,x)^i \vdash P ~ {\cal R} ~ Q$,
therefore $h.(x,x)^i \vdash P ~ {\cal R}_w ~ Q$. By Proposition~\ref{prop:soundness-bisim-upto},
this implies $h.(x,x)^i \vdash P ~ \obisim ~ Q.$
\qed
\end{proof}

\begin{lemma}
\label{lm:input-swap}
If $h_1.(x,x)^i.(y,y)^i.h_2 \vdash P ~ \obisim ~ Q$, where $x,y \not \in \freeFN{h_1,h_2}$,
then $h_1.(y,y)^i.(x,x)^i.h_2 \vdash P ~ \obisim ~ Q.$
\end{lemma}
\begin{proof}
We make use of soundness of the up-to techniques (Proposition~\ref{prop:soundness-bisim-upto}), 
more specifically, the up-to contraction and substitutions.
Note that a consequence of Proposition~\ref{prop:soundness-bisim-upto} is that 
$(\obisim)_t = \obisim$ for any $t \in \{\equiv,s,f,w,c,r,p \}.$
The applications of the up to techniques are as follows:
$$
\begin{array}{l}
h_1.(x,x)^i.(y,y)^i.h_2 \vdash P ~ \obisim ~ Q \\
\Downarrow \hbox{contraction, $x'$, $y'$ new names} \\
h_1.(y',y')^i.(x',x')^i.(x,x)^i.(y,y)^i.h_2 \vdash P ~ \obisim ~ Q\\
\Downarrow \hbox{substitution} \\
h_1.(y',y')^i.(x',x')^i.(x',x')^i.(y',y')^i.h_2 \vdash P[x'/x,y'/y] ~ \obisim ~ Q[x'/x,y'/y]\\
\Downarrow \hbox{weakening}\\
h_1.(y',y')^i.(x',x')^i.h_2 \vdash P[x'/x,y'/y] ~ \obisim ~ Q[x'/x,y'/y]\\
\Downarrow \hbox{substitution}\\
h_1.(y,y)^i.(x,x)^i.h_2 \vdash P ~ \obisim ~ Q
\end{array}
$$ 
\qed
\end{proof}

\begin{theorem}
The relation $\sim_o$ is a congruence on finite pure processes.
\end{theorem}
\begin{proof}
We show the relation $\sim_o$ are closed under all process contexts (except,
of course, replication).
It is enough to show closure under elementary context.

\begin{description}
\item[Input prefix] Suppose $P \sim_o Q$ and $x$ is a free name in $P$ and $Q$.
We show that $M(x).P \sim_o M(x).Q$ for all pure message $M.$
By definition, $h_1.(x,x)^i.h_2 \vdash P ~\obisim ~ Q$ for some bi-trace
$h_1.(x,x)^i.h_2.$ We assume that $h_1.h_2$ contains all the names
in $M$; otherwise apply the contraction rule to extend it to cover all the
names in $M$. This can be done because $\obisim$ is closed under bi-trace
extensions (Proposition ~\ref{prop:soundness-bisim-upto}). 
We then apply Lemma~\ref{lm:input-swap} to move the pair $(x,x)$ to the
end of the list. That is, we have 
$h_1.h_2.(x,x)^i \vdash P \sim_o Q.$ Note that since $M$ is an pure
message, by Lemma~\ref{lm:open-message}, $h_1.h_2\vdash M \eqm M.$
We can therefore apply Lemma~\ref{lm:clo-input}
to get $h_1.h_2 \vdash M(x).P \sim_o M(x).Q.$ 

\item[Output prefix] Suppose $P \sim_o Q$, i.e., $h \vdash P ~ \obisim ~ Q.$
We show that 
$h\vdash \spout M N P ~\obisim ~ \spout M N Q$, for any pure messages $M$ and $N$.
This amounts to showing that 
$h.(M,M)^i.(N,N)^o \vdash P ~\obisim ~ Q.$ 
This is indeed the case since $h.(M,M)^i.(N,N)^i \sqsubseteq_c h$ and 
$\obisim$ is closed under contraction of bi-traces. 

\item[Parallel composition] Suppose $h\vdash P ~ \obisim ~ Q.$
Let $R$ be any pure process. Then by Proposition~\ref{prop:soundness-bisim-upto},
$h' \vdash (P ~|~ R) ~ \obisim ~ (Q ~|~ R)$ for some universal trace $h'$
containing all the names of $P$, $Q$ and $R$. Therefore,
$(P~|~ R) ~\obisim ~ (Q ~ | ~ R).$ The left-composition, i.e., $(R ~ | ~ P) ~\sim_o (R~|~ Q)$
is proved analogously.

\item[Restriction] Suppose $P~\sim_o Q$, where 
$h_1.(x,x)^i.h_2 \vdash P ~\obisim ~ Q$.
We first use Lemma~\ref{lm:input-swap} to obtain $h_1.h_2.(x,x)^i \vdash P ~\obisim ~ Q$.
This is then followed by an up-to flexible-rigid reversal on $x$, weakening and finally the
restriction, to get $h_1.h_2 \vdash (\nu x)P ~\obisim ~ (\nu x)Q.$ 
Therefore, $(\nu x)P \sim_o (\nu x) Q.$

\item[Matching] In this case we first show the soundness of an up-to matching technique:
Given a consistent set of traced process pairs ${\cal R}$, define ${\cal R}_m$
the smallest set containing ${\cal R}$ and closed under the rule
$$
\infer[]
{h\vdash [M=N]P ~ {\cal R}_m [M = N]Q}
{h\vdash P ~ {\cal R} ~ Q, \hbox{ $M$ and $N$ are pure messages such that 
$\freeFN{M,N} \subseteq \freeFN{h}$}}
$$
and show that ${\cal R}_m$ is an open bisimulation whenever ${\cal R}$ is. 
This relies on the fact that,
for any consistent bi-trace $h$ and $h$-respectful substitution pair 
$\vec\theta = (\theta_1,\theta_2)$,
it holds that $h\vec\theta \vdash M\theta_1 \eqm M\theta_2$ and 
$h\vec\theta \vdash N\theta_1 \eqm N\theta_2$, and therefore by the consistency of $h\vec\theta$,
$M\theta_1 = N\theta_1$ if and only if $M\theta_2 = N\theta_2$. 
From this, it then follows that $(\obisim)_m = \obisim.$

We now show that $P \sim_o Q$ implies $[M = N] P \sim_o [M=N] Q$, for any pure
messages $M$ and $N.$ Suppose that $h\vdash P ~ \obisim ~ Q$.
Note that $M$ and $N$ may contain free names which are not free in $P$ and $Q$,
so we need to extend $h$ to a universal trace $h'$ containing all the names in $P$,
$Q$, $M$ and $N$. It would then follow that
$h'\vdash [M=N]P ~ {(\obisim)}_{cm} ~ [M=N] Q$, and therefore 
$[M=N]P \sim_o [M=N]Q.$

\item[Pairing] As in the previous case, we show that open bisimulation is closed
under the following rule: given a relation ${\cal R}$, 
define ${\cal R}_{l}$ to be the smallest relation containing
${\cal R}$ and closed under the rule
$$
\infer[]
{h \vdash (\splet x y M P) ~ {\cal R}_{l} ~ (\splet x y M Q)}
{h.(x,x)^i.(y,y)^i 
\vdash P ~ {\cal R} ~ Q, \hbox{ $x,y \not \in \freeFN{h}$, 
$M$ is an pure message and $\freeFN{M} \subseteq \freeNm{h}$}}
$$ 
We show that ${\cal R}_{l}$ is an open bisimulation up-to contraction,
given that ${\cal R}$ is an open bisimulation.
Let us examine one case here involving input action; the other two cases
can be handled similarly. 
Suppose 
$$h \vdash (\splet x y M P) ~ {\cal R}_{l} ~ (\splet x y M Q),$$
and $h.(x,x)^i.(y,y)^i \vdash P ~ {\cal R} ~ Q.$
Let $\vec\theta = (\theta_1,\theta_2)$ be a substitution pair respecting
$h$. We assume w.l.o.g. that $x \not \in \dom{\theta_1}.$
Suppose 
$$
\one{\splet x y {M\theta_1} {P\theta_1}}{U}{(z)P'}.
$$ 
It must be the case that 
$M\theta_1 = \spr {M_1} {M_2}$, $M\theta_2 = \spr {M_1'}{M_2'}$,
$h\vec\theta \vdash M_1 \eqm M_2$ and $h\vec\theta \vdash M_1' \eqm M_2'$
and $\one{P\theta_1[M_1/x,M_2/y]}{U}{(z)P'}.$  
Define the substitution pair  $\theta_1'$ and $\theta_2'$ as follows:
$$
\theta_1' = \theta_1 \cup \{M_1/x, M_2/y \}
\hbox{ and }
\theta_2' = \theta_2 \cup \{M_1'/x, M_2'/y \}.
$$ 
It is easy to see that $(\theta_1',\theta_2')$ respects $h.(x,x)^i.(y,y)^i$,
therefore we have
$\one{Q\theta_2'}{V}{(z)Q'}$ for some $V$ and $Q'$ such that 
$$
h\vec\theta.(M_1,M_1')^i.(M_2,M_2')^i.(U,V)^i.(z,z)^i \vdash P' ~ {\cal R}_l ~ Q'.
$$
Note that since $\freeFN{M} \subseteq \freeFN{h}$,
the free names of $M_1$, $M_2$, $M_1'$ and $M_2'$ are all in $h\vec\theta.$
We can therefore apply the weakening rule to the above traced process pair
to get
$$
h\vec\theta.(U,V)^i.(z,z)^i \vdash P' ~ ({\cal R}_{l})_c ~ Q'.
$$ 
Hence ${\cal R}_{l} \subseteq ~ \obisim$ by Proposition~\ref{prop:soundness-bisim-upto}.

Now we show that if $P \sim_o Q$ then 
$(\splet x y M P) \sim_o (\splet x y M Q)$ for any pure message $M$.
We can assume that $h.(x,x)^i.(y,y)^i \vdash P ~ \obisim ~ Q$ for
some universal trace $h$ (by applying
contraction and Lemma~\ref{lm:input-swap} to move the input pairs for $x$ and $y$),
and that $\freeFN{M} \subseteq \freeFN{h}.$ The latter means that $x$ and $y$ are
not in $\freeFN{M}.$ This is not a limitation since we can always apply renaming
to $x$ and $y$ in $P$ and $Q$ (recall that $\obisim$ is also closed under
respectful substitution) before we close it under the pairing context. 
Since $(\obisim)_l = \obisim$, we can apply the above closure rule and 
obtain 
$$h\vdash (\splet x y M P) ~ \obisim ~ (\splet x y M Q)$$
and therefore $(\splet x y M P) \sim_o (\splet x y M Q).$

\item[Encryption] This case is proved analogously to the case with pairing. In this case,
we define the closure under the case-expression: Let ${\cal R}$ be a relation. Then
${\cal R}_e$ is the smallest relation containing ${\cal R}$ and closed under the rule
$$
\infer[]
{h\vdash (\spcase M x N P) ~ {\cal R}_e ~ (\spcase M x N Q)}
{h.(x,x)^i \vdash P ~ {\cal R} ~ Q, 
\hbox{ $x\not \in \freeFN{h}$, $M$ and $N$ are pure messages and
$\freeFN{M,N} \subseteq \freeFN{h}$}}
$$
As in the previous case, we can show that ${\cal R}_e \subseteq ~ \obisim$, and therefore
$(\obisim)_e = ~ \obisim.$ The rest of the proof proceeds similarly to the previous case.
\end{description}
\qed
\end{proof}

\section{Conclusion and future work}
\label{sec:conc}

We have shown a formulation of open bisimulation for the spi-calculus. 
In this formulation, bisimulation is indexed by pairs of symbolic traces
that concisely encode the history of interactions between the environment
with the processes being checked for bisimilarity. 
We show that open bisimilarity is a congruence for finite processes
and is sound with respect to testing equivalence. 
For the latter, we note that with some minor modifications, we can also
show soundness of open bisimilarity with respect to barbed congruence.
Our formulation is directly inspired by hedged bisimulation~\cite{borgstrom05mscs}.
In fact, open bisimilarity can be shown to be sound with respect to hedged
bisimulation. Comparison with hedged bisimulation and other formulations of
bisimulation for the spi-calculus is left for future work. 

It would be interesting to see how the congruence results extend to 
the case with replications or recursions. This will probably require a more
general definition of the rule for up-to parallel composition. 
The definition of open bisimulation and the consistency of bi-traces make use of 
quantification over respectful substitutions. We will investigate whether there is a 
finite characterisation of  consistent bi-traces.
One possibility is to use a symbolic transition system, i.e., a transition
system parameterised upon certain logical constraints, the solution of which should
correspond to respectful substitutions. Some preliminary study in this direction
is done in \cite{briais07spi} for a variant of open bisimulation based on
hedged bisimulation. 
Since the bi-trace structure we use is a variant of symbolic traces, 
we will also investigate whether the techniques used for symbolic traces 
analysis~\cite{boreale01icalp} can be adapted to our setting.

Another interesting direction for future work is to find a {\em proof search}
encoding of the spi-calculus and open bisimulation in a logical framework.
This has been done for open bisimulation for the $\pi$-calculus~\cite{tiu04fguc},
in a logical framework based on intuitionistic logic~\cite{miller05tocl}. 
The logic used in that formalization features a new quantifier, called $\nabla$,
which allows one to reason about ``freshness'' of names, a feature crucial
to the correct formalization of the notion of name restriction in the $\pi$-calculus.
An interesting aspect of this formalization is the fact that quantifier alternation
in logic, i.e., the alternation between universal quantifer and $\nabla$, captures
a certain natural class of name-distinctions. 
Adapted to our definition of open bisimulation, it would seem that rigid names
should be interpreted as $\nabla$ quantified names, whereas non-rigid names should
be interpreted universally quantified names. Details of such a proof search encoding
for the spi-calculus are left for future work.

\paragraph{Acknowledgment}
This paper is a revised and extended version of a conference version presented at APLAS 2007 \cite{tiu07aplas}.
The author thanks the anonymous referees for their comments on an earlier draft of
the conference version of the paper. Jeremy Dawson has formalized in Isabelle/HOL\footnote{The proof scripts are available
on \url{http://users.rsise.anu.edu.au/~jeremy/isabelle/2005/spi/}} most of the results in Section~\ref{sec:obsv} 
concerning observer theories and some results in Section~\ref{sec:open} concerning properties of bi-traces.
He has also given many useful comments. This work is supported by
the Australian Research Council, under Discovery Project DP0880549.

\end{document}